# Insulating Electronic States Near the Dirac Point Arising from Twisted Stacking and Curvature in 3D Nanoporous Graphene

*Yoichi Tanabe[1]\*, Hayato Sueyoshi[2], Samuel Jeong[2], Kojiro Imai[1], Shojiro Kimura[3], Yoshikazu Ito[2,4]\**

[1]Department of Applied Science, Faculty of Science, Okayama University of Science, Okayama 700-0005, Japan

[2]Department of Applied Physics, Institute of Pure and Applied Sciences, University of Tsukuba, 1-1-1 Tennodai, Tsukuba, Ibaraki, 305-8573, Japan

[3]Institute for Materials Research, Tohoku University, Sendai 980-8577, Japan

[4]Tsukuba Institute for Advanced Research (TIAR), University of Tsukuba, 1-1-1 Tennodai, Tsukuba, Ibaraki, 305-8577, Japan

**Corresponding Author**

Yoichi Tanabe; Email: tanabe@ous.ac.jp

Yoshikazu Ito; Email: ito.yoshikazu.ga@u.tsukuba.ac.jp

Abbreviations: (3D-NPG: three-dimensional nanoporous graphene, EDLT: electric double-layer transistor, CVD: chemical vapor deposition, DEME-TFSI: *N*,*N*-Diethyl-2-methoxy-*N*-methylethan-1-aminium Bis[(trifluoromethyl)sulfonyl]amide, 2D graphene: two-dimensional graphene)

Abstract: Twist-stacked graphene with a twist angle $\theta$ of $\sim 5° - 30°$ retains two-dimensional monolayer graphene–like Dirac states near the Dirac point. In three-dimensional nanoporous graphene (3D-NPG), curvature inherently produces twist-stacking and topological defects required to form a porous network. When regions with $\theta \geq 5°$ dominate, Dirac states in individual layers are expected to persist, allowing the Dirac-electron behavior to be tuned through coupling to the 3D curved geometry. However, predicted band gap formation or localized states have remained unobserved. Here we report that 3D-NPG maintains monolayer-like Dirac electronic states while simultaneously exhibiting insulating behavior near the Dirac point. Raman G-band softening confirms these monolayer-like states, and an Arrhenius-type temperature–resistance trend coexisting with weak localization near the Dirac point indicates partially insulating states induced by topological defects. These findings demonstrate that 3D-NPG hosts distinctive Dirac electronic states coupled to 3D curvature, providing a platform for developing new functionalities in 3D graphene-based electronics and energy devices.

**Graphical abstract**

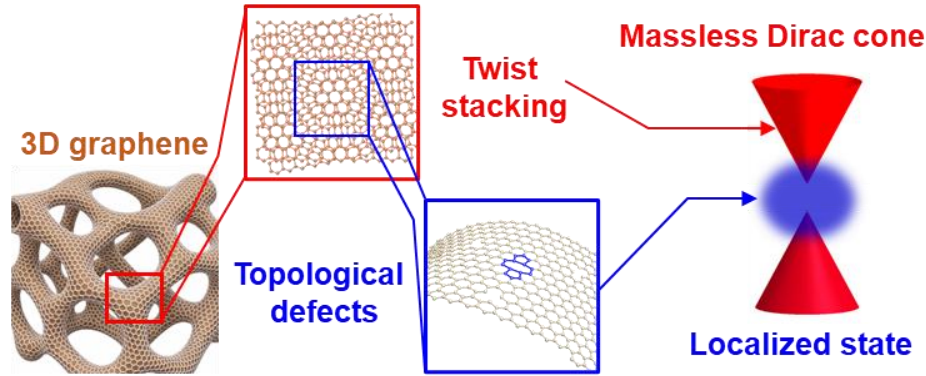

## 1. Introduction

A two-dimensional (2D) monolayer graphene sheet, which is an atomic layer of carbon arranged in a honeycomb lattice, hosts massless Dirac electronic states and functions as a versatile platform for Dirac-electron functionalities[1,2]. To expand these functionalities and enable broader applications, two-dimensional graphene has recently been engineered into three-dimensional graphene using various methods[3–13]. These studies have identified two key geometric features of 3D nanoarchitectures that are central to tailoring Dirac-electron behavior: curved surfaces containing topological defects and their stacking configurations (Fig. 1a), which enable the tuning of electronic and phonon properties through geometry-driven interactions.

Curvature engineering via topological defects enables the realization of nonplanar and curved graphene nanoarchitectures[5–13]. A monolithic, bi-continuous 3D nanoporous graphene (3D-NPG) network geometrically requires topological defects on its curved surfaces to form the tubular architecture[10]. On these curved surfaces (Fig. 1a), lattice mismatch arising from curvature and defect formation promotes twist stacking, resulting in domains with random twist angles (moiré domains)[14].

In these moiré domains, the interlayer twist angle $\theta$ in few-layer graphene serves as a key parameter for tuning electronic states. In twisted bilayer graphene, the moiré superlattice with period $L = \frac{a}{2\sin\frac{\theta}{2}}$ (where $a$ is the graphene lattice constant) generates alternating AA and AB domains and a pseudo periodicity that leads to moiré miniband formation[3,15]. At the magic angle $\theta \sim 1.1°$, where the long-period moiré pattern $L \sim 13$ nm brings Dirac cones from the K and K' points into proximity within the moiré Brillouin zone, interlayer tunneling produces flat bands near the Dirac points[16], enabling strongly correlated states such as superconductivity[17] and Mott insulating behavior[18]. In contrast, at $\theta \gtrsim 5°$ with $L \leq 2.8$ nm, the enlarged moiré Brillouin zone separates the Dirac cones, interlayer coupling nearly vanishes near the Dirac point, and each layer retains monolayer-like massless Dirac cone states within at least ±0.5 eV (e.g., 0.5 eV at $\theta = 5°$, Figure 1a) as estimated using the twist-angle dependent flat-band energy with respect to the Dirac point $E_{VHS} = \hbar v \frac{k_\theta}{2} = \hbar v K \sin\left(\frac{\theta}{2}\right)$ ($K = \frac{4\pi}{3a}, k_\theta = 2K \sin\left(\frac{\theta}{2}\right)$)[4,19]. This behavior contrasts with AB-stacked bilayer graphene, in which strong interlayer coupling transforms the Dirac cone into a parabolic band[20]. Thus, the twist angle can be used as a control parameter that switches between strongly correlated Dirac electrons and ordinary massless Dirac electrons in multilayer graphene.

Thus, for samples with sufficiently large twist angle distributions (i.e., $\theta \geq 5°$), Dirac electrons near the Dirac points are largely decoupled between the multilayers[4]. Consequently, the electronic properties of 3D nanoarchitectures near the charge neutrality points can be interpreted within a monolayer-graphene framework, in which geometry-induced perturbations provide the primary mechanism for modifying the Dirac spectrum. Curvature-driven defects and strain can induce nontrivial inter- and intravalley scattering[21] and may lead to curvature-induced Landau quantization[22], band-gap opening[23,24], or broadening of the Dirac point into localized electronic states[25], often accompanied by a strain-induced phonon softening[26].

To probe these Dirac electronic states, in situ Raman spectroscopy under field-effect carrier doping using an electric double-layer transistor (EDLT) is a powerful method. In two-dimensional graphene, the G band associated with the in-plane $E_{2g}$ phonon at the $\Gamma$ point exhibits softening, i.e., a decrease in the frequency $\omega_G$ as the Fermi level $E_F$ approaches the Dirac point due to nonadiabatic electron-phonon coupling[27–29]. The resulting $E_F - \omega_G$ relation allows identification of monolayer-like Dirac states in various three-dimensional graphene systems, and $E_F$ estimated from capacitance-based carrier density has been compared with $E_F$ extracted from this Raman-derived curve[30]. In planar twisted bilayer graphene with a twist angle of $\sim11°$, using a top-gated EDLT configuration asymmetrically tuned the carrier densities of the top and bottom layers[31]. This asymmetric doping induced different $E_F$ offsets from the Dirac point in each layer, producing distinct self-energy corrections of the G-mode phonon and resulting in G-band splitting. The observed G-band softening due to asymmetric carrier doping indicates that both layers retain monolayer-like Dirac electronic states. In 3D-NPG, an EDLT formed by immersing a freestanding 3D-NPG structure in an ionic liquid induces asymmetric doping between the outermost and interior layers (Fig. 1b), producing a Fermi-level ($E_F$) offset that enables layer-selective spectroscopic observation of Dirac states. Such layer-selective spectroscopy on 3D-NPG is expected to suggest unique Dirac states in 3D graphene that integrate curved surfaces and twist-stacked layers, yet a comprehensive study examining their interplay is still lacking.

In this study, we elucidated intrinsic Dirac electronic properties near the Dirac points of 3D-NPG, which consists of curved surfaces and twist-stacked layers formed within a monolithic, bi-continuous nanoporous structure, by combining in situ Raman spectroscopy with electrical transport measurements using EDLTs. Identifying Dirac electronic states in individual curved layers through Raman analysis enables a clearer interpretation of how intrinsic Dirac behavior couples to the three-dimensional geometry in electrical transport measurements. This integrated approach provides a platform for exploring novel Dirac-electron properties emerging in 3D graphene nanoarchitectures and for advancing functionalities in three-dimensional graphene devices[32,33].

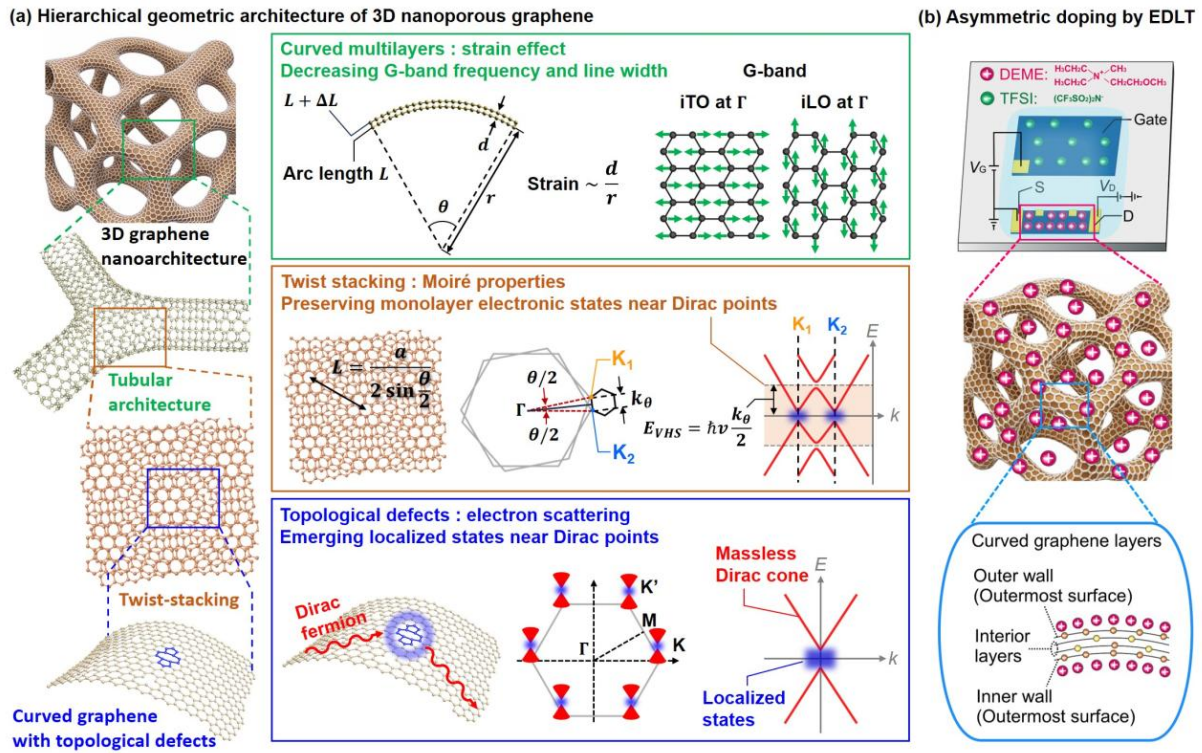

**Figure 1. Structure and electronic properties of 3D graphene nanoarchitecture.** (a) Hierarchical geometric structure of the 3D graphene nanoarchitecture and the associated physical properties. The architecture can be hierarchically decomposed into tubular networks, curved multilayers with randomly twisted stacking, and curved monolayer graphene containing

topological defects. The tubular geometry is expected to induce a downshift of the G-band phonon frequency due to strain. In the randomly twisted stacking, where twist angles are predominantly ~5° or greater, interlayer hybridization near the Dirac point is suppressed, and the electronic structure of each layer retains monolayer graphene-like character. Furthermore, scattering associated with topological defects can lead to localized states near the Dirac point and the emergence of a transport gap. (b) Asymmetric doping of curved multilayer graphene in 3D nanoporous graphene using an electric-double-layer transistor (EDLT).

## 2. Materials and methods

3D-NPG with a curvature radius of 100–150 nm was synthesized using a nanoporous nickel-based CVD method[10] (see Supplementary Material, Sections 1.1 and 1.2). SEM images (Fig. S1) showed a monolithic three-dimensional graphene network with 100–150 nm pores and bi-continuously connected cross-sections. The TEM image (Fig. 2a) revealed an interconnected tubular graphene network with a curvature radius of ~100 nm. High-resolution TEM images at the edge confirmed that the graphene consists of bilayer structures (Fig. 2b), and the inset shows a selected-area electron diffraction pattern characteristic of high-quality few-layer graphene. Statistical analysis of layer numbers performed across several TEM images at different positions indicated an average thickness of $2.8 \pm 1.3$ layers (Fig. S2).

HR-TEM imagesrevealed several moiré patterns with periods $L$ ranging from ~0.5 nm to ~1.7 nm at different positions within the same 3D-NPG sample (Figs. 2c–e). The patterns at ~1.1 nm and ~1.7 nm (Figs. 2c and 2d) showed clear periodicity, whereas the ~0.5 nm pattern (Fig. 2e) displayed nonuniform periodic features (see Supplementary Material, Section 2.1 and Fig. S3). The twist angles corresponding to ~1.1 nm and ~1.7 nm were estimated to be $\sim 13°$ and $\sim 8°$, respectively. In contrast, the ~0.5 nm pattern corresponded to a twist angle of ~28°, and its lack of translational periodicity suggests a quasicrystal-like motif typical of twisted bilayer graphene near $\theta \sim 30°$[34]. The coexistence of these distinct periodicities indicates that the curved surfaces host moiré domains spanning a wide range of twist angles. We also examined whether curvature-dependent moiré patterns could modify moiré-related physical properties. HR-TEM observations on 3D-NPG samples with curvature radii of 25–50 nm, 100–150 nm, and 500–1000 nm revealed no discernible changes in moiré periodicity (see Supplementary Material, Section 2.1 and Fig. S4), consistent with previous photoemission spectroscopy results,[21] which showed a linear density of states as a function of binding energy and no pronounced density of states peak near the Dirac point. Therefore, the 3D curved surfaces primarily generate randomly oriented moiré domains rather than systematically curvature-controlled twist angles.

Raman spectroscopy and X-ray photoelectron spectroscopy (XPS) were used to characterize the graphene features. The Raman spectra showed the characteristic D ( $\sim 1350$ cm$^{-1}$ ), G ($\sim 1580$ cm$^{-1}$), and 2D ($\sim 2690$ cm$^{-1}$) bands (Fig. 2f). The intensity ratio of 2D and G bands ($I_{2D}/I_G$) was 1.7, being consistent with few-layer graphene, and the intensity ratio of D and G bands ($I_D/I_G$) was 0.12, indicating a defect density comparable to that reported for 2D CVD graphene[35,36]. The XPS C 1s spectrum displayed a sharp C–C peak at ~284 eV (FWHM ~0.7 eV) and a small amount (~6%) of oxygen-containing groups, including epoxy C–O, carbonyl C=O, and carboxyl COO in the 286–290 eV range (Fig. 2g)[10], consistent with CVD-grown graphene[36]. These results confirm that 3D-NPG retains high-quality few-layer graphene characteristics. Residual Ni from the CVD template was below the XPS detection limit (Fig. S5).

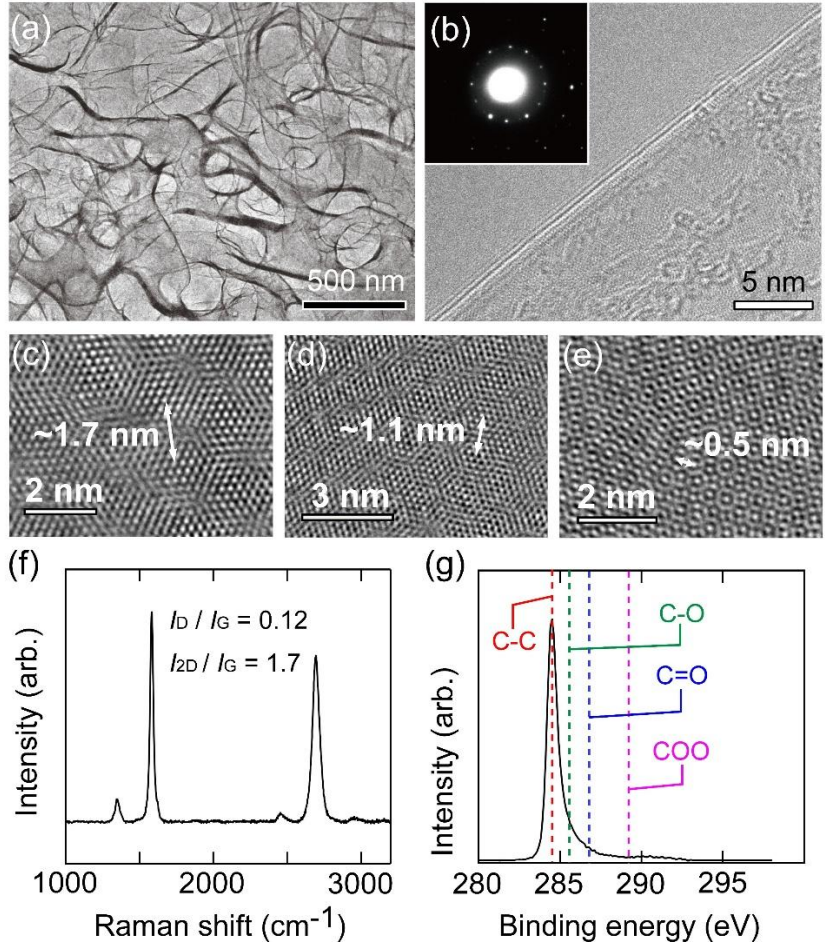

**Figure 2. Structural and chemical properties of 3D nanoporous graphene.** (a) A low-magnification STEM image. (b) TEM image at the edge and electron diffraction patterns. (c-e) A HR-TEM image. The arrows indicate the characteristic scale of the moiré pattern. (f) Raman spectra and (g) XPS of 3D nanoporous graphene.

3D-NPG-based EDLTs using the ionic liquid DEME-TFSI[37] were fabricated for in situ Raman and electrical transport measurements (see Supplementary Material, Section 1.3). To resolve electronic states near the Dirac point, residual contamination and charged impurities must be minimized (see Supplementary Material, Section 1.3.1), as they induce carrier scattering and spatially inhomogeneous doping that obscure intrinsic Dirac-electron behavior. Two devices were prepared: Device 1 used as-prepared 3D-NPG repeatedly washed with water, and Device 2 used 3D-NPG electrochemically etched in ionic liquid under an applied gate voltage ($V_G = -3$ V for 1 h at 240 K) (see Supplementary Material, Section 1.3.1). A two-dimensional bilayer graphene sample was also fabricated for comparison. The temperature and magnetic field dependence of the electrical resistance was measured using a 4-probe method, while the $E_F$ was tuned toward the Dirac point via EDL gating. This combined approach, electrochemical channel etching together with EDL-gated fine control of $E_F$, represents the first attempt in 3D-NPG to monitor electronic transport properties in the vicinity of the Dirac point. Thus, 3D-NPG with different curvature radii were also examined. For the NPG with smallest pores (25–50 nm), the network was mechanically fragile due to the high density of topological defects, making it difficult to acquire electrical transport data from low temperatures up to near room temperature without structural failure under ionic-liquid infiltration. For the NPG with larger pores (500–1000 nm, Fig. S6), similar Raman measurements were conducted. Electrostatic screening in the thicker multilayers induces a strong carrier-density imbalance between the outermost and inner layers, preventing the layer-averaged $E_F$ from being tuned sufficiently close to the Dirac point. Therefore, intrinsic Dirac-point transport could not be reliably extracted (see Supplementary Material, Section 2.2 and Fig. S7).

## 3. Results

Figure 3a shows Raman spectra of Device 1 acquired without intensity normalization at $V_G = -1.545$ V (p-type), 0.045 V (Dirac-point region), and 1.545 V (n-type doping) (see full $V_G$ range in Fig. S8). The G-band intensity increased near the Dirac point region ($V_G = 0.045$ V), but the strong resonance typically observed for a single twist angle $\theta \sim 11°$ did not appear under 532 nm excitation (Fig. S9) due to randomly distributed twist angles (Supplementary Material, Section 2.3)[31]. The G-band splitting, linewidth broadening, and blue shift were evident at $V_G = -1.545$ V while peak splitting was absent and broadening weaker at $V_G = 1.545$ V. This asymmetry between p- and n-type doping can be attributed to adiabatic effects, whereby increasing electron doping across the Dirac point leads to a monotonic decrease in G-band frequency[38]. An $R_{moire}$ band associated with the moiré reciprocal-lattice vector[39] was not clearly observed because of the random twist-angle distribution. The D-band full width at half maximum (FWHM) broadened from 31 to 60 cm$^{-1}$ across the $V_G$ range, while the $I_D/I_G$ ratio (~0.1) remained low from $-0.945$ to 1.245 V (Fig. S10), indicating that mechanisms beyond defect-activated intervalley scattering contribute to D-band formation. This behavior suggests a moiré-assisted double resonance, in which moiré reciprocal-lattice vectors provide the required momentum in combination with long-range impurity scattering and electron-phonon coupling[39–41].

All G bands across the full $V_G$ range in Device 1 were fit with single or double Lorentzian functions to determine the $V_G$ dependence of their peak position ($\omega_G$) and FWHM ($\Gamma_G$) (Fig. 3b, c, Fig. S8), with results summarized in Fig. 3d, e. In the range of $-0.045 \leq V_G \leq 0.225$ V near the Dirac point, the electrical resistance reached its maximum (Fig. S11), while $\omega_G$ reached a minimum of 1581 cm$^{-1}$ and $\Gamma_G$ showed a local maximum of 24 cm$^{-1}$. When positive (0.225 V $\leq$ $V_G \leq 0.765$ V) or negative ($-0.225$ V $\leq V_G \leq -0.045$ V) gate voltages injected electrons or holes, $\omega_G$ increased and $\Gamma_G$ decreased. These behaviors reflect nonadiabatic electron-phonon coupling near the Dirac point, where coupling between the G-mode phonon and interband electron–hole excitations induces self-energy renormalization and reduces the phonon lifetime [27–29]. In the p-type region in $-0.585$ V $< V_G \leq -0.225$ V, $\Gamma_G$ increased from $\sim 23$ cm$^{-1}$ to $\sim 26$ cm$^{-1}$ and $\omega_G$ shifted from 1585 cm$^{-1}$ to 1587 cm$^{-1}$ with decreasing $V_G$. In particular, in the p-type doping region in $V_G \leq -0.585$ V, the G-bands split into two distinct peaks. Conversely, in the n-type region $V_G \geq 0.765$ V , $\Gamma_G$ increased from ~22 cm$^{-1}$ to ~26 cm$^{-1}$ without peak splitting, accompanied by a blue shift of $\omega_G$ from ~1581 cm$^{-1}$ to ~1587 cm$^{-1}$.

The observed G-band broadening and splitting under carrier doping ($V_G \leq -0.315$ V and $V_G \geq 0.765$ V) in 3D-NPG arise from layer-dependent carrier densities across non-planar, curved, twist-stacked few-layer graphene domains (Fig. 1b). In curved bilayer graphene on 3D-NPG, nearly symmetric charging occurs on the two outermost layers because no interior layers exist (Supplementary Material, Section 2.4, Fig. S12). In contrast, in curved few-layer graphene (layer number $\geq$ 3; Fig. 1b), increasing interior layers create a charge imbalance between outer surfaces and inner layers. This imbalance produces distinct G-band blue shifts across layers, manifesting as G-band broadening ($V_G \leq -0.315$ V and $V_G \geq 0.765$ V, Fig. 3d, e) and eventually G-band splitting ($V_G \leq -0.585$ V; Fig. 3d, e). Similar G-band splitting arising from asymmetric layer doping has been reported in top-gate planar twisted-bilayer graphene EDLTs[31].

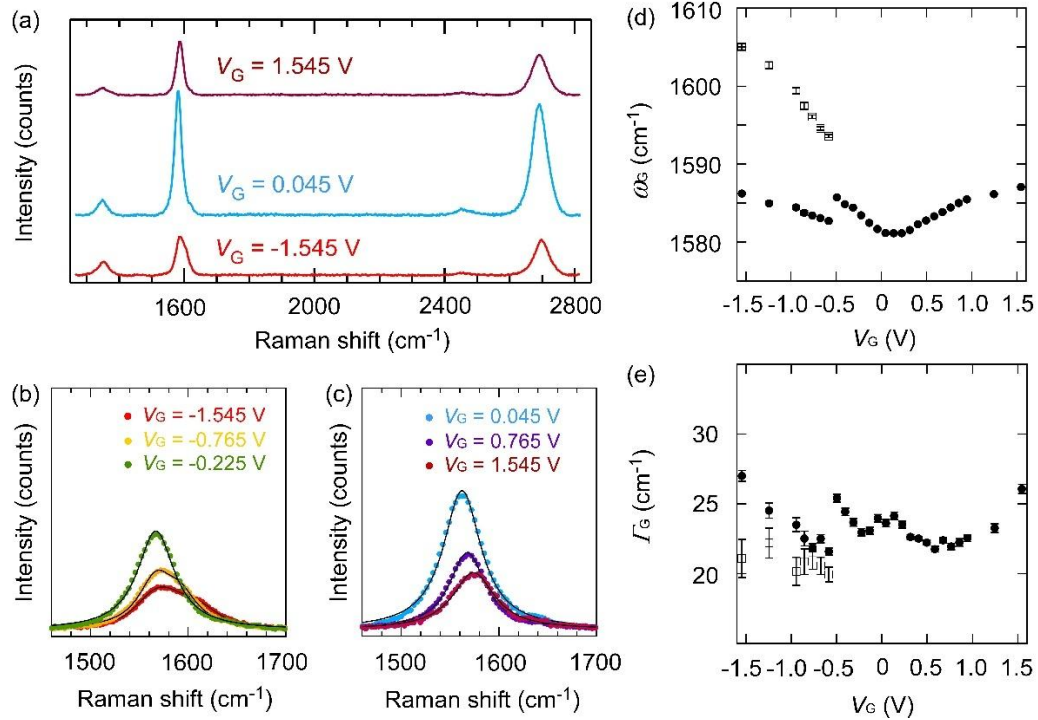

**Figure 3. Raman spectra of 3D nanoporous graphene on EDLTs using Device 1.** (a) Raman spectra obtained after subtracting the background signal from the ionic liquid gel, at $V_G = -1.545$ V (p-type region), 0.045 V (near the Dirac point), and 1.545 V (n-type region). During the measurements, the position, accumulation time, and laser power were fixed. (b, c) Raman G-band at $-1.545 \leq V_G \leq 1.545$ V. Solid lines represent Lorentzian fitting curves. $V_G$ dependence of (d) G-band frequency, $\omega_G$, (e) full width at half maximum $\Gamma_G$ over the full $V_G$ range.

We quantitatively analyzed the G-band softening in 3D-NPG using a model established for 2D graphene to validate the $V_G$ dependent Fermi level $E_F$. First, we applied the $E_F$-dependent phonon self-energy correction for 2D graphene, $|E_\mathrm{F}| \times 42 = \omega_\mathrm{G} - \omega_\mathrm{G0}$ [31]. Substituting $\omega_G$ into this relation with $\omega_{G0} = 1578$ cm$^{-1}$, obtained from the $V_G - \omega_G$ curve outside the splitting regime, yielded an averaged $E_F$. Next, we calculated the carrier density $n$ by integrating the EDL capacitance over $V_G$, using the average layer number ($2.8 \pm 1.3$), and then estimated the averaged $E_F$ from $E_F = \hbar v_F \sqrt{\pi |n|}$ for 2D graphene (Supplementary Material, Section 2.5, Fig. S13). The $V_G$ dependence of both $E_F$ values showed good agreement (Fig. 4), confirming that G-band shifts in 3D-NPG follow the nonadiabatic electron-phonon coupling expected for 2D graphene within this $V_G$ range.

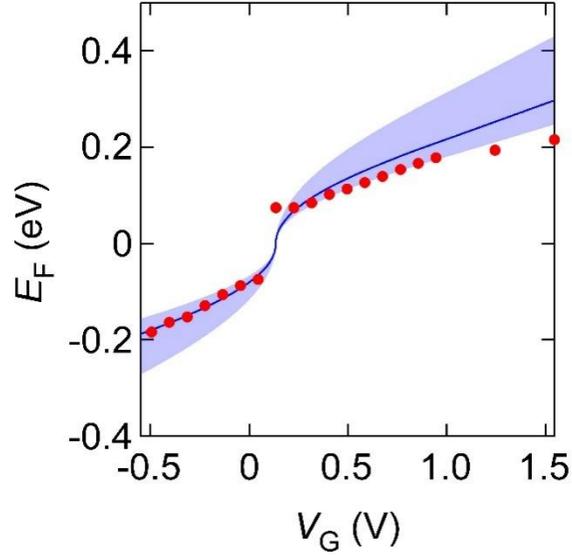

**Figure 4. Comparison of $E_F$ estimated from Raman G-band and capacitance measurements**. $V_G$ dependence of averaged Fermi levels ($E_F$) (red circles), estimated from the G-band frequency $\omega_G$ using $|E_F| \times 42 = \omega_G - \omega_{G0}$, excluding the $V_G$ range corresponding to the G-band splitting regime. For comparison, the averaged $E_F$ calculated from the $V_G$ dependence of the EDL capacitance (blue line) using the average layer number (2.8±1.3) is also plotted. The blue shaded region represents the uncertainty in $E_F$ arising from the error in the average layer number (See Supplementary Material, Section 2.5).

After confirming that the G-band shifts follow the nonadiabatic electron-phonon coupling of 2D graphene, we estimated the electron-phonon coupling strength $\lambda_\Gamma$ from the $E_F$ dependence of $\omega_G$ and $\Gamma_G$ near the Dirac points (Figure S14a, b). We chose the fitting range where no G-band broadening from asymmetric doping of outermost and interior layers appeared (Supplementary Material, Section 2.6). As shown in Fig. S14a, b, theoretical self-energy curves for 2D graphene, convolved with a Gaussian energy-fluctuation width $\delta E_F$ to account for electron–hole puddles, reproduced the experimental data. The calculated $\lambda_\Gamma$ was 0.03, close to the value reported for 2D graphene[28,30], confirming that the electron–phonon coupling strength in 3D-NPG resembles that of 2D graphene. The fitted baseline parameters ($\Gamma_G^0 = 19$ cm$^{-1}$ and $\omega_G^0 = 1578$ cm$^{-1}$) differed from those of twisted-bilayer graphene with an 11° twist ( $\Gamma_G^0 = 8$ cm$^{-1}$ and $\omega_G^0 = 1583$ cm$^{-1}$ )[31], where Dirac electronic states are also preserved but are not influenced by curvature-induced strain. For a curved trilayer graphene with curvature radius $\rho \sim 100$ nm and thickness $t \sim 0.67$ nm, the resulting top-to-bottom strain gradient is $\sim 0.67$ %. Based on the strain dependence of $\omega_G$ and $\Gamma_G$ in 2D graphene[26], an average layer strain of 0.34 % is expected to shift $\omega_G^0$ by ~3–10 cm$^{-1}$, while a 0.67 % interlayer strain difference is expected to increase $\Gamma_G^0$ by ~5–20 cm$^{-1}$. Thus, the layer-dependent Raman G-band softening observed in 3D-NPG is consistent with the intrinsic G-band behavior of single-layer graphene.

Raman G-band analysis showed that the electronic states near the Dirac point in 3D-NPG match those of 2D graphene while incorporating strain arising from 3D curvature. These results indicated that, due to randomly oriented moiré domains with twist angles of 8° or larger, as identified by HR-TEM, interlayer hybridization between the curved multilayer graphene is suppressed within an energy window $\gtrsim 1.5$ eV centered around the Dirac point. Consequently, in 3D-NPG, electronic states intrinsic to "3D curved monolayer graphene" emerged. These states are derived from the Dirac-electron states of monolayer graphene, while incorporating strain effects arising from curvature and scattering effects originating from topological defects. Thus, the Dirac-electron behavior coupled to 3D geometric features was further investigated through electrical transport measurements.

Device 1 was first examined; however, transport near the Dirac point was strongly influenced by charged-impurity scattering, so the electrochemically etched Device 2 was used instead. The electrical resistance ($R$) of Device 2 near the Dirac point was measured to probe curvature-induced

Dirac-electron behavior (Supplementary Material, Section 2.7 and Figs. S11, S15, and S16). At 240 K, it exhibited ambipolar transport with a pronounced $R(V_G)$ peak (Fig. S11). We applied $V_G$ = −0.47 V to tune $E_F$ near the Dirac point and measured the temperature ($T$) dependence of the resistance (Fig. 5a). The $T$-$R$ curve exhibited a negative temperature coefficient. The resistance increased logarithmically below 20 K and decreased with increasing temperature at higher $T$ in a concave-down, Arrhenius-like manner, dropping by nearly 50 % between 1.4 K and 150 K. Such Arrhenius-like $T$-$R$ behavior has not been observed in previous 3D-NPG electrical transport studies[21] due to the absence of electrochemical etching and constitutes the first experimental observation capturing the intrinsic nature of charge transport in the vicinity of the Dirac point.

To interpret the negative $T$-$R$ curves near the Dirac point, we considered thermally excited carriers from puddles created by local charged-impurity doping[42,43]. With a Fermi-energy fluctuation $\delta E$, the conductivity follows $\sigma(T) = \sigma_0 [1 + \sqrt{\frac{2}{\pi}}(k_B T/\delta E) + (\pi^2/3)(k_B T/\delta E)^2]$. Estimating the electron–hole puddle density from the FWHM of the resistance peak[42] ($\delta n \sim 10^{12}$ cm$^{-2}$, Fig. S17), we obtained $\delta E \sim 0.12$ eV. This predicts only an ~11% conductivity increase between 1.4 K and 150 K, far smaller than the nearly twofold experimental rise, indicating that puddle-driven thermally excited carriers only partially explain the behavior.

An Arrhenius plot of $R(T)$ showed that between 80 K and 110 K the resistance followed an $\exp(\varDelta/2k_B T)$ dependence (Fig. 5b), where $\varDelta$ is the transport gap, indicating band-gap formation or nearest-neighbor hopping[44,45]. In typical semiconductors at low $T$, a crossover in activation energy reduces the $R(T)$ slope or drives a transition to variable-range hopping. In contrast, our devices showed a $\ln T$ dependence of $R(T)$ (Fig. 5c), characteristic of weak localization and consistent with the weak-localization features in magnetoresistance (Supplementary Material, Section 2.7, Fig. S15b)[21,46]. These results suggest partially insulating electronic states in 3D-NPG that hinder itinerant Dirac-electron transport through the graphene labyrinth. In other words, this intrinsic electrical transport can be readily masked by residual impurity-induced local doping and by a substantial offset of $E_F$ from the Dirac point[21]. Thus, directly observing an exponential *T-R* dependence under properly controlled experimental conditions is essential to establish gapped insulating behavior in the itinerant Dirac-electron system (see also Supplementary Material, Section 2.8 and Fig. S18).

To capture this behavior, we fitted the temperature-dependent conductivity $G$ near the Dirac point using a phenomenological model :

$$G(T) = A \ln T + B \left[ 1 + \sqrt{\frac{2}{\pi}} \left( \frac{k_B T}{\delta E} \right) + \left( \frac{\pi^2}{3} \right) \left( \frac{k_B T}{\delta E} \right)^2 \right] + C \exp\left( -\frac{\Delta}{2 k_B T} \right) \quad (1).$$

This model reproduced the insulating $G(T)$ curves and yielded transport gaps $\Delta$ of $\sim 48$ meV (Fig. 5d), higher than two-dimensional CVD graphene but lower than graphene oxide (0.11–0.13 eV) [36] (Table S1).

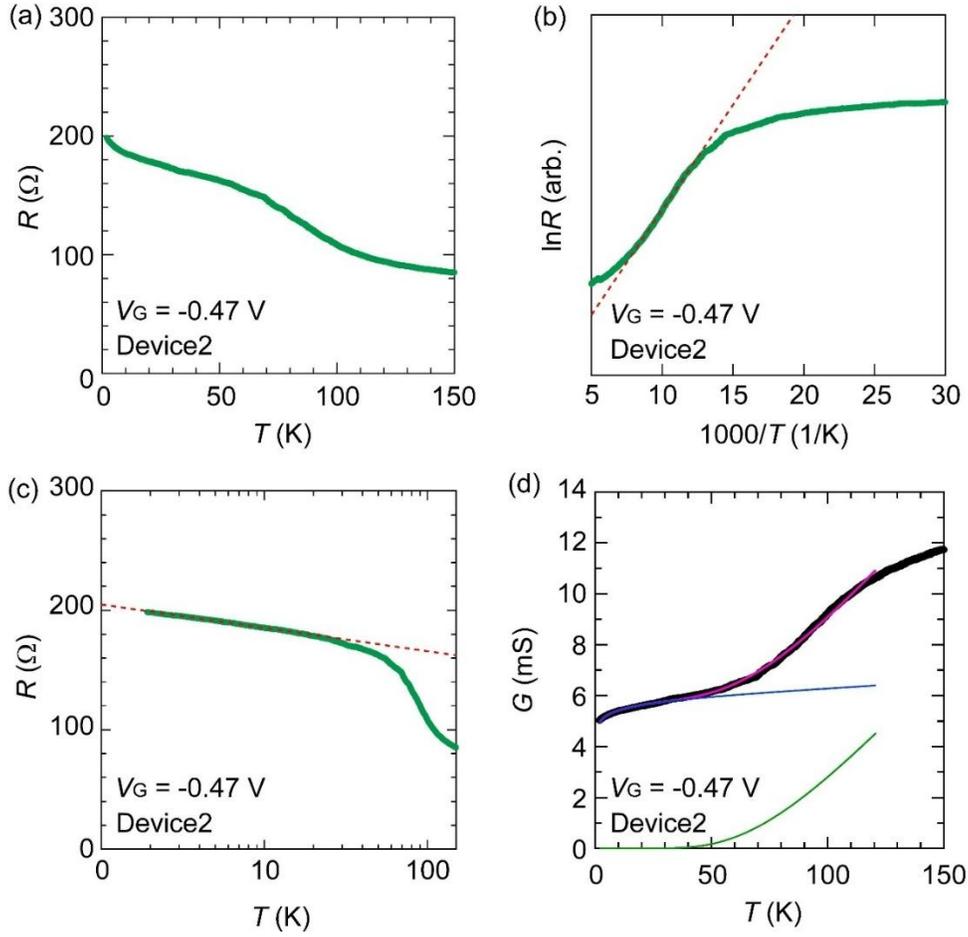

**Figure 5. Temperature dependence of the electrical resistance of 3D nanoporous graphene EDLTs using Device 2.** (a) $T - R$ curve at $V_G = -0.47$ V (near the Dirac point). (b) Arrhenius plot of $1000/T$ vs. $\ln R$. The red dashed line showed a linear fit. (c) $\ln T - R$ plot. The red dashed lines show the results of the $\ln T - R$ fitting. (d) Temperature dependence of the conductance $G$ near the Dirac point. Red curves were fitted with Eq. (1); blue curves represent the weak-localization contribution and the thermally excited contribution from electron–hole puddles; green curves represent the Arrhenius-type thermally activated term from insulating electronic states.

## 4. Discussion

We next examined the origin of the insulating states near the Dirac points. First, we evaluated whether the Arrhenius-type $R(T)$ behavior could arise from graphene oxidation. Table S1 summarizes the Raman $I_D/I_G$ ratios, XPS-derived oxygen-to-carbon (O/C) ratios, and the transport gaps $\Delta$ from the $T$-$R$ analysis. $I_D/I_G$ ratio of Device 2 is lower than that of zero-gap CVD graphene[36]. In addition, the O/C ratio of Device 2 remained well below that of graphene oxide[36], and the increase in the O/C ratio after electrochemical etching originates from electron-beam-induced oxidation during the XPS measurements (Table S1, Supplementary Material, Section 2.9 and Fig. S19). Therefore, oxidation did not significantly contribute to the Arrhenius-type $R(T)$ behavior in Device 2.

Next, we evaluated whether a pseudomagnetic field induced by lattice expansion and contraction in non-planar, curved graphene could explain the transport behavior. Within small-curvature Dirac theory, the scalar curvature $X = 2K$ ($K$: Gaussian curvature) modifies the energy spectrum as $E(p) = \hbar v_F \sqrt{p^2 + \frac{X}{12}}$, producing a curvature-induced gap, $\Delta_{curve} = 2\hbar v_F \sqrt{\frac{X}{12}}$ on dome-shaped regions ($X > 0$).[24] For a curvature radius of $\rho = 100$ nm ($X = 2/\rho^2$), the resulting gap is $\Delta_{curve} \sim 5 - 6$ meV, nearly an order of magnitude smaller than the experimental transport gap $\Delta \sim 48$ meV. Thus, curvature alone cannot account for the observed Arrhenius-type behavior.

We also considered whether finite-size quantization in tubular segments of the 3D network, analogous to carbon nanotubes, could contribute to the Arrhenius-type $R(T)$ behavior. The characteristic energy scales can be estimated as $E_d = \frac{2t_0 r_0}{d}$ for diameter quantization[47] and $E_L = \frac{\hbar v_F}{2L}$ for length quantization[48], where $t_0$ is the nearest-neighbor hopping, $r_0$ is the C–C bond length, $d$ is the tube diameter, and $L$ is the tube length. Using the geometrical scales in the TEM images (Fig.2(a)) (i.e., $d \sim 100$–150 nm, $L \sim 200$–400 nm), we obtain $E_d \sim 3$–4 meV and $E_L \sim 1$–2 meV, which is one order of magnitude smaller than the experimental transport gap. Thus, finite-size effects alone cannot account for the observed Arrhenius-type behavior (Supplementary Material, Section 2.10.).

We examined whether moiré domain boundaries could be responsible for the insulating behavior. However, unlike previous reports on well-ordered boundaries[49], we did not resolve clear, extended domain boundaries in our HR-TEM observations. This absence may be explained by the effective release of geometric frustration in 3D-NPG through distributed topological defects in curved regions, which prevents significant lattice mismatch. Moreover, the pronounced reduction in resistance and the emergence of a more graphene-like magnetoresistance after electrochemical etching indicate that charged-impurity scattering, rather than domain-boundary scattering, is the dominant limitation of transport. Any contributions from local boundaries are therefore expected to be minor (Supplementary Material, Section 2.11.).

Finally, we evaluated whether geometric configuration and scattering from topological defects—such as 5–8–5 ring defects ($\sim$0.1 at. % per carbon)—could generate localized electronic states. These defects are predicted to broaden the Dirac point into $\sim$0.05 eV–wide localized states[25]. Because such defects were identified in our HR-TEM images (Fig. S20), the thermally activated resistance is likely associated with nearest-neighbor hopping through these localized states. Therefore, scattering arising from the various topological defects required to form the curved surface architecture may underlie the insulating behavior observed near the Dirac point.

To further assess whether such defect-induced gap formation is global or local, we provide an order-of-magnitude estimate of the curvature-required non-hexagonal rings based on the Gauss–Bonnet theorem[50]. For a representative radius of curvature of 100 nm, the inferred defect fraction is extremely dilute. Therefore, it is more natural to interpret the transport gap arising from curvature-required defects as forming locally in limited regions of the 3D labyrinth, rather than uniformly across the entire structure (Supplementary Material, Section 2.12). We also note that curvature-based counting is intrinsically insensitive to net-zero charge defects such as 5–8–5 structures and may underestimate their presence. Together with the HR-TEM observations and the coexistence of insulating and itinerant channels, this supports predominantly local gap formation (Supplementary Material, Section 2.12.).

## 5.  Conclusions

In conclusion, we investigated a monolithic, bi-continuous 3D nanoporous graphene architecture composed of non-planar, curved, twist-stacked few-layer domains and found that it preserves 2D-like Dirac electronic states near the Dirac point through combined in situ Raman and electrical-transport measurements using electric double-layer transistors. Interactions between Dirac electrons and topological defects within the curved geometry can locally induce insulating electronic states. Raman measurements revealed that asymmetric doping between the outermost and interior layers enables layer-dependent G-band shifts and G-band splitting. These behaviors are well described by phonon self-energy corrections arising from nonadiabatic electron-phonon coupling in 2D graphene, together with curvature-induced strain effects. Electrical transport measurements showed pronounced weak localization and a $\ln T$ dependence of $R(T)$ coexisting with Arrhenius behavior, indicating partially insulating states with a ~48 meV transport gap near the Dirac point. Topological defects such as 5–8–5 rings likely generate localized states responsible for this gap. Thus, non-planar, curved, twist-stacked graphene incorporating topological defects offers promising routes for tuning electronic states and enabling new functionalities in 3D graphene devices.

**Acknowledgement**

We would like to thank Prof. Yasuhiro Hatsugai, Prof. Tomonari Mizoguchi, and Prof. Kazuki Sone for fruitful discussions. We are also grateful to Ms. Kazuyo Omura at the Institute for Material Research, Tohoku University for the XPS measurements. This work was funded by JSPS-Kakenhi (Grant Numbers JP22K04867, JP23K17661, JP24H00478), Tokuyama Science Foundation, Iwatani Naoji Foundation and CASIO Science Promotion Foundation. This work was performed at the Cooperative Research and Development Center for Advanced Materials (CRDAM, proposal no. 202412-CRKEQ-0009) and the High Field Laboratory for Superconducting Materials (HFLSM, proposals no. 202311-HMKPA-0005, 202211-HMKPA-0005, and 202111-HMKPA-0002) under the GIMRT program of the Institute for Materials Research at Tohoku University. It was also performed at the Research Instruments Center (MPMS-XL5, Quantum Design) at Okayama University of Science.

**Author Statement**

**Yoichi Tanabe**: Conceptualization, Methodology, Validation, Formal analysis, Investigation, Resources, Data Curation, Writing - Original Draft, Writing - Review & Editing, Visualization, Project administration, Funding acquisition. **Hayato Sueyoshi**: Validation, Formal analysis, Investigation, Resources, Data Curation. **Samuel Jeong**: Investigation. **Kojiro Imai**: Investigation, Data Curation. **Shojiro Kimura**: Investigation. **Yoshikazu Ito**: Conceptualization, Methodology, Resources, Data Curation, Writing - Original Draft, Writing - Review & Editing, Project administration, Funding acquisition

**References**

[1] A.K. Geim, K.S. Novoselov, The rise of graphene, Nat. Mater. 6 (2007) 183–191. https://doi.org/10.1038/nmat1849

[2] A.H. Castro Neto, F. Guinea, N.M.R. Peres, K.S. Novoselov, A.K. Geim, The electronic properties of graphene, Rev. Mod. Phys. 81 (2009) 109–162. https://doi.org/10.1103/RevModPhys.81.109

[3] R. Bistritzer, A.H. MacDonald, Moiré bands in twisted double-layer graphene, Proc. Natl. Acad. Sci. U.S.A. 108 (2011) 12233–12237. https://doi.org/10.1073/pnas.1108174108

[4] A. Luican, G. Li, A. Reina, J. Kong, R.R. Nair, K.S. Novoselov, A.K. Geim, E.Y. Andrei, Single-layer behavior and its breakdown in twisted graphene layers, Phys. Rev. Lett. 106 (2011) 126802. https://doi.org/10.1103/PhysRevLett.106.126802

[5] L. Wu, S. Qiu, Y. Guo, X. Geng, M. Chen, S. Liao, C. Zhu, Y. Gong, M. Long, J. Xu, X. Wei, M. Sun, L. Liu, High-density three-dimension graphene macroscopic objects for high-capacity removal of heavy metal ions, Sci. Rep. 3 (2013) 2125. https://doi.org/10.1038/srep02125

[6] Z.-S. Wu, Y. Sun, Y.-Z. Tan, S. Yang, X. Feng, K. Müllen, Three-Dimensional Graphene-Based Macro- and Mesoporous Frameworks for High-Performance Electrochemical Capacitive Energy Storage, J. Am. Chem. Soc. 134 (2012) 19532–19535. https://doi.org/10.1021/ja308676h

[7] Y. Xu, Z. Lin, X. Huang, Y. Liu, Y. Huang, X. Duan, Flexible solid-state supercapacitors based on three-dimensional graphene hydrogel films, ACS Nano 7 (2013) 4042–4049. https://doi.org/10.1021/nn4000836

[8] Z. Chen, W. Ren, L. Gao, B. Liu, S. Pei, H.-M. Cheng, Three-dimensional flexible and conductive interconnected graphene networks grown by chemical vapor deposition, Nat. Mater. 10 (2011) 424–428. https://doi.org/10.1038/nmat3001

[9] X. Cao, Y. Shi, W. Shi, G. Lu, X. Huang, Q. Yan, Q. Zhang, H. Zhang, Preparation of novel 3D graphene networks for supercapacitors applications, Small 7 (2011) 3163–3168. https://doi.org/10.1002/smll.201100990

[10] Y. Ito, Y. Tanabe, H.-J. Qiu, K. Sugawara, S. Heguri, N.H. Tu, K.K. Huynh, T. Fujita, T. Takahashi, K. Tanigaki, M. Chen, High-quality three-dimensional nanoporous graphene, Angew. Chem. Int. Ed. 53 (2014) 4822–4826.

[11] K. Nomura, H. Nishihara, N. Kobayashi, T. Asada, Kyotani, 4.4 V supercapacitors based on super-stable mesoporous carbon sheet made of edge-free graphene walls, Energy Environ. Sci. 12 (2019) 1542–1549. https://doi.org/10.1039/C8EE03184C

[12] Y. Ito, Y. Tanabe, K. Sugawara, M. Koshino, T. Takahashi, K. Tanigaki, H. Aoki, M. Chen, Three-dimensional porous graphene networks expand graphene-based electronic device applications, Phys. Chem. Chem. Phys. 20 (2018) 6024–6033. https://doi.org/10.1039/C7CP07667C

[13] I. Di Bernardo, G. Avvisati, C. Mariani, N. Motta, C. Chen, J. Avila, M.C. Asensio, S. Lupi, Y. Ito, M. Chen, T. Fujita, M.G. Betti, Two-dimensional hallmark of highly interconnected three-

dimensional nanoporous graphene, ACS Omega 2 (2017) 3691–3697. https://doi.org/10.1021/acsomega.7b00706

[14] O. De Luca, A. Palamara, M. Pisarra, F. Mazzei, T. Caruso, G. Desiderio, D. Marchiani, R. Frisenda, S. Jeong, Y. Ito, C. Mariani, M.G. Betti, M. Papagno, R.G. Agostino, D. Pacilé, A. Sindona, Imaging of twisted monolayers in three-dimensional nanoporous graphene, Phys. Rev. B 111 (2025) 045432. https://doi.org/10.1103/PhysRevB.111.045432

[15] N.N.T. Nam, M. Koshino, Lattice relaxation and energy band modulation in twisted bilayer graphene, Phys. Rev. B 96 (2017) 075311.

[16] A. Kerelsky, L.J. McGilly, D.M. Kennes, L. Xian, M. Yankowitz, S. Chen, K. Watanabe, T. Taniguchi, J. Hone, C. Dean, A. Rubio, A.N. Pasupathy, Maximized electron interactions at the magic angle in twisted bilayer graphene, Nature 572 (2019) 95–100. https://doi.org/10.1038/s41586-019-1431-9

[17] Y. Cao, V. Fatemi, S. Fang, K. Watanabe, T. Taniguchi, E. Kaxiras, P. Jarillo-Herrero, Unconventional superconductivity in magic-angle graphene superlattices, Nature 556 (2018) 43–50. https://doi.org/10.1038/nature26160

[18] Y. Cao, V. Fatemi, A. Demir, S. Fang, S.L. Tomarken, J.Y. Luo, J.D. Sanchez-Yamagishi, K. Watanabe, T. Taniguchi, E. Kaxiras, R.C. Ashoori, P. Jarillo-Herrero, Correlated insulator behaviour at half-filling in magic-angle graphene superlattices, Nature 556 (2018) 80–84. https://doi.org/10.1038/nature26154

[19] T. Ohta, J.T. Robinson, P.J. Feibelman, A. Bostwick, E. Rotenberg, T.E. Beechem, Evidence for Interlayer Coupling and Moiré Periodic Potentials in Twisted Bilayer Graphene, Phys. Rev. Lett. 109 (2012) 186807. https://doi.org/10.1103/PhysRevLett.109.186807

[20] E. McCann, Asymmetry gap in the electronic band structure of bilayer graphene, Phys. Rev. B 74 (2006) 161403(R). https://doi.org/10.1103/PhysRevB.74.161403

[21] Y. Tanabe, Y. Ito, K. Sugawara, M. Koshino, S. Kimura, T. Naito, I. Johnson, T. Takahashi, M. Chen, Dirac Fermion Kinetics in 3D Curved Graphene, Adv. Mater. 32 (2020) 2005838. https://doi.org/10.1002/adma.202005838

[22] F. Guinea, M.I. Katsnelson, A.K. Geim, Energy gaps, topological insulator state and zero-field quantum Hall effect in graphene by strain engineering, Nat. Phys. 6 (2010) 30–33. https://doi.org/10.1038/nphys1420

[23] J. da Silva-Araújo, H. Chacham, R.W. Nunes, Gap opening in topological-defect lattices in graphene, Phys. Rev. B 81 (2010) 193405. https://doi.org/10.1103/PhysRevB.81.193405

[24] P. Castro-Villarreal, R. Ruiz-Sánchez, Pseudomagnetic field in curved graphene, Phys. Rev. B 95 (2017) 125432. https://doi.org/10.1103/PhysRevB.95.125432

[25] P. Kot, J. Parnell, S. Habibian, C. Straßer, P.M. Ostrovsky, C.R. Ast, Band dispersion of graphene with structural defects, Phys. Rev. B 101 (2020) 235116. https://doi.org/10.1103/PhysRevB.101.235116

[26] T.M.G. Mohiuddin, A. Lombardo, R.R. Nair, A. Bonetti, G. Savini, R. Jalil, N. Bonini, D.M. Basko, C. Galiotis, N. Marzari, K.S. Novoselov, A.K. Geim, A.C. Ferrari, Uniaxial strain in

graphene by Raman spectroscopy: G peak splitting, Grüneisen parameters, and sample orientation, Phys. Rev. B 79 (2009) 205433. https://doi.org/10.1103/PhysRevB.79.205433

[27] A. Das, S. Pisana, B. Chakraborty, et al., Monitoring dopants by Raman scattering in an electrochemically top-gated graphene transistor, Nat. Nanotechnol. 3 (2008) 210–215. https://doi.org/10.1038/nnano.2008.67

[28] S. Pisana, M. Lazzeri, C. Casiraghi, et al., Breakdown of the adiabatic Born–Oppenheimer approximation in graphene, Nat. Mater. 6 (2007) 198–201. https://doi.org/10.1038/nmat1846

[29] J. Yan, Y. Zhang, P. Kim, A. Pinczuk, Electric field effect tuning of electron–phonon coupling in graphene, Phys. Rev. Lett. 98 (2007) 166802. https://doi.org/10.1103/PhysRevLett.98.166802

[30] G. Froehlicher, S. Berciaud, Raman spectroscopy of electrochemically gated graphene transistors: Geometrical capacitance, electron–phonon, electron–electron, and electron–defect scattering, Phys. Rev. B 91 (2015) 205413. https://doi.org/10.1103/PhysRevB.91.205413

[31] T.F. Chung, R. He, T.L. Wu, Y.P. Chen, Optical phonons in twisted bilayer graphene with gate induced asymmetric doping, Nano Lett. 15 (2015) 1203–1210. https://doi.org/10.1021/nl504318a

[32] A. Dechant, T. Ohto, Y. Ito, M.V. Makarova, Y. Kawabe, T. Agari, H. Kumai, Y. Takahashi, H. Naito, M. Kotani, Geometric model of 3D curved graphene with chemical dopants, Carbon 182 (2021) 223–232. https://doi.org/10.1016/j.carbon.2021.06.004

[33] Y. Ito, Y. Shen, D. Hojo, Y. Itagaki, T. Fujita, L. Chen, T. Aida, Z. Tang, T. Adschiri, M. Chen, Correlation between chemical dopants and topological defects in catalytically active

nanoporous graphene, Adv. Mater. 28 (2016) 10644–10651. https://doi.org/10.1002/adma.201604318

[34] P. Moon, M. Koshino, Y.-W. Son, Quasicrystalline electronic states in 30° rotated twisted bilayer graphene, Phys. Rev. B 99 (2019) 165430. https://doi.org/10.1103/PhysRevB.99.165430

[35] A.C. Ferrari, D.M. Basko, Raman spectroscopy as a versatile tool for studying the properties of graphene, Nat. Nanotechnol. 8 (2013) 235–246. https://doi.org/10.1038/nnano.2013.46

[36] A.I. Aria, A.W. Gani, M. Gharib, Effect of dry oxidation on the energy gap and chemical composition of CVD graphene on nickel, Appl. Surf. Sci. 293 (2014) 1–11. https://doi.org/10.1016/j.apsusc.2013.11.117

[37] Y. Tanabe, Y. Ito, K. Sugawara, D. Hojo, M. Koshino, T. Fujita, T. Aida, X. Xu, K.K. Huynh, H. Shimotani, T. Adschiri, T. Takahashi, K. Tanigaki, H. Aoki, M. Chen, Electric Properties of Dirac Fermions Captured into 3D Nanoporous Graphene Networks, Adv. Mater. 28 (2016) 10304–10310. https://doi.org/10.1002/adma.201601067

[38] M. Lazzeri, F. Mauri, Nonadiabatic Kohn anomaly in a doped graphene monolayer, Phys. Rev. Lett. 97 (2006) 266407. https://doi.org/10.1103/PhysRevLett.97.266407

[39] A. Righi, S.D. Costa, H. Chacham, C. Fantini, P. Venezuela, C. Magnuson, L. Colombo, W.S. Bacsa, R.S. Ruoff, M.A. Pimenta, Graphene Moiré patterns observed by umklapp double resonance Raman scattering, Phys. Rev. B 84 (2011) 241409(R). https://doi.org/10.1103/PhysRevB.84.241409

[40] L.G. Cancado, A. Jorio, E.H. Martins Ferreira, F. Stavale, C.A. Achete, R.B. Capaz, M.V.O. Moutinho, A. Lombardo, T.S. Kulmala, A.C. Ferrari, Quantifying defects in graphene via Raman spectroscopy at different excitation energies, Nano Lett. 11 (2011) 3190–3196. https://doi.org/10.1021/nl201432g

[41] V. Carozo, C.M. Almeida, B. Fragneaud, P.M. Bedê, M.V.O. Moutinho, J. Ribeiro-Soares, N.F. Andrade, A.G. Souza Filho, M.J.S. Matos, B. Wang, M. Terrones, R.B. Capaz, A. Jorio, C.A. Achete, L.G. Cançado, Resonance effects on the Raman spectra of graphene superlattices, Phys. Rev. B 88 (2013) 085401. https://doi.org/10.1103/PhysRevB.88.085401

[42] Y.-W. Tan, Y. Zhang, K. Bolotin, Y. Zhao, S. Adam, E.H. Hwang, S. Das Sarma, H.L. Stormer, P. Kim, Measurement of Scattering Rate and Minimum Conductivity in Graphene, Phys. Rev. Lett. 99 (2007) 246803. https://doi.org/10.1103/PhysRevLett.99.246803

[43] Q. Li, E.H. Hwang, S. Das Sarma, Disorder-induced temperature-dependent transport in graphene, Phys. Rev. B 84 (2011) 115442. https://doi.org/10.1103/PhysRevB.84.115442

[44] C. Kittel, Introduction to Solid State Physics, 8th ed., Wiley, Hoboken, NJ, 2005.

[45] N.F. Mott, Conduction in Non-Crystalline Materials, Philos. Mag. 19 (1969) 835–852. https://doi.org/10.1080/14786436908216338

[46] E. McCann, K. Kechedzhi, V.I. Fal'ko, H. Suzuura, T. Ando, B.L. Altshuler, Weak-localization magnetoresistance and valley symmetry in graphene, Phys. Rev. Lett. 97 (2006) 146805. https://doi.org/10.1103/PhysRevLett.97.146805

[47] C.T. White, J.W. Mintmire, Density of states reflects diameter in nanotubes, Nature 394 (1998) 29–30. https://doi.org/10.1038/27801

[48] S.G. Lemay, J.W. Janssen, M. van den Hout, M. Mooij, M.J. Bronikowski, P.A. Willis, R.E. Smalley, L.P. Kouwenhoven, C. Dekker, Two-dimensional imaging of electronic wavefunctions in carbon nanotubes, Nature **412** (2001) 617–620. https://doi.org/10.1038/35088013

[49] T. Ma, Z. Liu, J. Wen, Y. Gao, X. Ren, H. Chen, C. Jin, X.-L. Ma, N. Xu, H.-M. Cheng, W. Ren, Tailoring the thermal and electrical transport properties of graphene films by grain size engineering, Nat. Commun. **8** (2017) 14486. https://doi.org/10.1038/ncomms14486

[50] M. Hayashi, Differential geometry and morphology of graphitic carbon materials, Phys. Lett. A **342** (2005) 237–246. https://doi.org/10.1016/j.physleta.2005.05.037

# Supplementary Material of
# Insulating Electronic States Near the Dirac Point Arising from Twisted Stacking and Curvature in 3D Nanoporous Graphene

*Yoichi Tanabe[1]\*, Hayato Sueyoshi[2], Samuel Jeong[2], Kojiro Imai[1], Shojiro Kimura[3], Yoshikazu Ito[2,4]\**

[1]Department of Applied Science, Faculty of Science, Okayama University of Science, Okayama 700-0005, Japan

[2] Department of Applied Physics, Institute of Pure and Applied Sciences, University of Tsukuba, 1-1-1 Tennodai, Tsukuba, Ibaraki, 305-8573, Japan

[3] Institute for Materials Research, Tohoku University, Sendai 980-8577, Japan

[4]Tsukuba Institute for Advanced Research (TIAR), University of Tsukuba, 1-1-1 Tennodai, Tsukuba, Ibaraki, 305-8577, Japan

## 1. Materials and methods

## 1.1 Materials preparation

### 1.1.1 2D bilayer graphene

Bilayer 2D graphene on a Cu foil was purchased by AirMembrane Corporation. The graphene was exfoliated as following previous reports (for example, Science, 2009, 324, 1312-1314).

### 1.1.2. Preparation of nanoporous Ni CVD substrate

$Ni_{30}Mn_{70}$ ingots were prepared by melting pure Ni and Mn (purity >99.9 at.%) using an Ar-protected arc melting furnace. After annealing at 900°C for 24 hours for microstructure and composition homogenization, the ingots were cold-rolled to thin sheets with a thickness of ~50 μm at room temperature. Nanoporous Ni was prepared by chemical dealloying in a 1.0 M $(NH_4)_2SO_4$ (purity $\geq$ 99.5%, Wako) aqueous solution at 50°C. After dealloying, the samples were rinsed thoroughly with water and ethanol and dried in vacuum.

### 1.1.3. Preparation of the various nanoporous graphene

Nanoporous Ni substrates loaded in a quartz tube (φ26×φ22×250 mm) were inserted into the center of a quartz tube (φ30×φ27×1000 mm) furnace and annealed at 800°C (G800) or 950°C (G950) under 200 sccm Ar (purity $\geq$ 99.99%) and 100 sccm $H_2$ (purity $\geq$ 99.99%) for 3 mins or 120 min, respectively. After the reduction pre-treatment, benzene (0.5 mbar, 99.8%, anhydrous, Sigma-Aldrich) was introduced with the gas flow of Ar (200 sccm) and $H_2$ (100 sccm) at 800°C or 950°C for graphene growth for 2 min or

5 sec. The furnace was immediately opened, and the inner quartz tube was cooled with a fan to room temperature. The graphene-grown nanoporous Ni substrates were dissolved by 1.0 M HCl (assay 35～37% w/w, Wako) solution overnight and the resulting graphene samples were then transferred into a 2.0 M HCl solution to remove residual Ni completely. The samples were repeatedly washed with distilled water and kept in 2-propanol (IPA, 99.7 wt%, FUJIFILM Wako Chemical Corp.) solution for a supercritical $CO_2$ fluid drying to perform structural characterizations and measurements.

### 1.1.4. Procedure of the supercritical $CO_2$ drying method

To prevent collapse and damage of the 3D curved structure of graphene by the capillary force of water during drying, the graphene samples were immersed in IPA and dried using supercritical $CO_2$ (sc$CO_2$) to gradually substitute IPA with minimized capillary force. The samples were first transferred to a glass bottle (volume: 5 mL) filled with IPA (400 μL), and then the bottle was placed in an 80 mL pressure-resistant container (TAIATSU techno Corp). After removing the air inside the container with gaseous $CO_2$ purging, the pressure of the container gradually increased to 15 MPa by introducing liquid $CO_2$ (5 MPa, −4°C, the density of 0.964 g/mL) at a flow rate of 20 mL/min (19 g/min) using a high-pressure plunger pump (NIHON SEIMITSU KAGAKU Co. Ltd, NP-KX-540). The sc$CO_2$ drying process was carried out at 70°C with constant $CO_2$ flowing at a rate of 5 mL/min (4.8 g/min) by forming a homogeneous phase of IPA and sc$CO_2$ to minimize the capillary force. The pressure was maintained at 15 MPa during the drying procedure for at least 5 h. After drying completely, the temperature was set at 40°C, and it gradually depressurized over 43 h from 15 MPa to atmospheric pressure by gradually venting $CO_2$ from the system.

## 1.2. Sample characterizations

The morphology and microstructure of the graphene samples were characterized through scanning electron microscopy (SEM, JEOL JCM-7000 NeoScope), and transmission electron microscopy (TEM; Hitachi, HF5000). The samples were transferred on a Cu grid without a carbon support film. Chemical analyses of the graphene samples were conducted by using an X-ray Photoelectron Spectroscopy (XPS, AXIS ultra DLD, Shimazu) with Al $K$a using X-ray monochromator. The samples were pasted on a conductive carbon tape. The XPS peaks were fitted according to CasaXPS software. Raman spectra were recorded by using a micro-Raman spectrometer (Renishaw InVia Reflex 532) with an incident wavelength of 532.5 nm. The laser power was set at 0.1 mW to avoid possible damage or unexpected reduction by laser irradiation. The graphene samples were placed on a background-free glass slide. The accumulation time of each spectrum is 100 s.

## 1.3. Device fabrication and measurements

### 1.3.1. Fabrications of electric double layer transistor

3D graphene electric double-layer transistors (EDLTs) using the ionic liquid DEME-TFSI were fabricated. Two types of transistor devices were prepared: Device 1 (as-prepared 3D-NPG with repeatably water washing process) and Device 2 (electrochemically washing using EDLTs at gate voltages ($V_G = -3$ V for 1 h at 240 K) after the supercritical $CO_2$ drying. Device 1 and Device 2 were fabricated using different growth batches under same CVD experimental conditions. For Raman measurements of Device 1, to improve stability under ambient conditions during measurements, we dissolved DEME-TFSI (KANTO CHEMICAL Co., Inc., 99.96%) and Poly(vinylidene fluoride-*co*-hexafluoropropylene) (PVDF-HFP) (Sigma Aldrich) in acetone at a mass ratio of 4:1, then allowed the acetone to evaporate to gel the ionic liquid [S1]. Transport measurements were performed using either DEME-TFSI or a DEME-TFSI/PVDF gel. Electrical resistance was measured in a nitrogen gas-filled glovebox or in a $^4$He refrigerator.

### 1.3.2. Raman and electrical transport measurements

Raman spectra of 3D nanoporous graphene EDLT were measured using Device 1 at room temperature in the range of gate voltage ($V_G$) from –1.5 V to +1.5 V. At each voltage step, the Raman spectrum of the sample were sequentially measured and, in parallel, the background spectrum of the ionic-liquid gel was also measured. Subtracting the Raman spectrum of gel background from the Raman spectrum of samples yielded the intrinsic Raman response of the graphene. The semiconductor parameter analyzer (Agilent B1500) was used to apply the gate voltages to the transistor device.

Electrical transport measurements were carried out using four probe methods using Cu wires. The four probe resistances of the devices were measured using a NF 5660/5650/5645 lock-in amplifiers operated at a frequency of 13 Hz. The Keithley 2400 or ADCMT 6242 source meter was used to apply DC gate voltages in the $V_G$ range of $-1.5$ V to 1.5 V. The $V_G - R$ curve measurements were carried out inside a nitrogen-filled grove box at 240 K using Peltier device. Moreover, the low-temperature transport measurements were performed under the constant $V_G$ output using helium-4 refrigerator in the temperature ranges of 1.4 K to 240 K under magnetic fields of $-18$ T to 18 T (18T-SM, JASTEC) and in the 1.8 K to 240 K under magnetic fields of $-5$ T to 5 T (MPMS-

XL5, Quantum Design).

### 1.3.3 Capacitance measurements

The capacitance was measured using the LCR meter (IM3533, HIOKI) in the $V_G$ range of $-1.5$ V to 1.5 V and in the AC voltage frequency range of 0.1 Hz to 100,000 Hz at room temperature.

## 2. Supporting Discussion

### 2.1 Interpretations of Moiré patterns

HR-TEM image of 3D nanoporous graphene was shown in Fig. S3. A modulation of the moiré length scale was observed in the range $L = \sim 1 - 1.5$ nm corresponding to $\theta \sim 9° - 14°$. Moreover, in Fig. S3, a $\sim 3$ nm contrast modulation which was longer than the primary moiré periods ($1 - 1.5$ nm; $\theta \sim 9° - 14°$) was also observed. This is consistent with a moiré-of-moiré superlattice and a suggestion of twisted stacking involving three or more layers [S2].

To estimate the energy ranges preserving the 2D graphene-like Dirac electronic states, the twisted angle is used to estimate the van Hove singularity energy (i.e., flat band energy). The van Hove singularity energy relative to the Dirac point in twisted bilayer graphene was given by $E_{VHS} = \hbar v K \sin\left(\frac{\theta}{2}\right)$, where $K = 4\pi/3a$ [S3]. Similar $E_{VHS}$ shift with increasing twist angles was also predicted in the twisted trilayer graphene case [S4]. Since the minimum twist angle of $\theta \sim 8°$ corresponds to $E_{VHS} \sim 0.8$ eV, the moiré domains on 3D-NPG are expected to preserve 2D graphene-like Dirac dispersion within the energy window of approximately $\pm 0.8$ eV relative to the Dirac points, corresponding to carrier densities up to $\sim 4 \times 10^{13}$ cm$^{-2}$. Beyond this energy window, the flat-band density-of-states (DOS) contributions from each moiré domain can be superimposed on the 2D graphene-like linear DOS. However, because in 3D-NPG twist angles were randomly distributed across moiré domains, causing the flat bands to emerge at different $E_{VHS}$ values. Thus, the small volume fraction of any individual domain prevents pronounced flat band peaks in the total DOS. Consequently, the linear DOS could be preserved in 3D-

NPG, consistent with photoemission spectroscopy [S5].

We also observed moiré patterns of 3D-NPG with curvature radius of 25-50 nm [S11], and 500-1000 nm with high resolution TEM measurements. We revealed that no discernible curvature-radius-dependent changes in the moiré periodicity (Figure S4). Moreover, photoemission spectroscopy (PES) measurements reported in previous reports [S11] showed that no pronounced density-of-states peak near the Dirac point was observed on these samples, indicating that the twist angle near the magic angle can be negligible. Therefore, the curvature radius does not affect the moiré angles and curved morphology bring the random moiré angles on the 3D-NPG.

## 2. 2. EDLT measurements for 3D-NPG with curvature radius of 25-50 nm and 500-1000 nm

To clarify the curvature radius dependence of low temperature transport behaviors, we also examined the EDLT measurements for 3D-NPG with curvature radius of 25-50 nm and 500-1000 nm reported in Ref. S11. For the 25–50 nm case, the network is mechanically fragile due to the high density of topological defects, making it impossible to acquire electrical transport data from low temperature up to near room temperature without structural failure under ionic-liquid infiltration. For 500–1000 nm case, to check the tuning of $E_F$ by gating, we carried out the in situ Raman measurements as shown in Figure S7. Our Raman analysis of the $V_G$ dependence of the G band indicates that gate-induced modulation is expected primarily in the outermost layers on 500-1000 nm graphene with 4.5±3.9 layers (Figure S6), charge injection into inner layers is strongly suppressed due to electrostatic screening effects, yielding only a very weak change in the layer-averaged carrier density. Accordingly, we observed almost no gate-voltage-induced shift of the G band $\omega_G$ and $\Gamma_G$.

Based on the Raman discussion, compared with a sample with a curvature radius of 100-150 nm shown in the main text, a sample with a curvature radius of 500-1000 nm, which has roughly twice as many layers, shows a larger difference in carrier concentration between the outermost layer and the inner layers when carrier control is performed using an EDLT, due to the electrostatic screening effect. Therefore, even if a gate voltage is applied to tune the carrier concentration to the vicinity of the charge-neutrality point and transport is measured as in the 100-150 nm sample, the carrier-concentration inhomogeneity is large, and the intrinsic behavior near the charge-neutrality point cannot be extracted.

## 2. 3. Evaluating the enhancement of the G-band intensity resulting from resonance between the 532 nm excitation laser and the flat-band energy

In localized regions of the sample with an approximate 11° twist angle, we examined whether the 532 nm laser photon energy resonates with $E_{VHS}$, leading to an enhancement of the G band intensity near the Dirac point. In monolayer graphene, electron–phonon coupling reduced the intensity ratio of $I_G/I_{2D}$ by ~50% [S6]. In twisted bilayer graphene with an 11° twist angle, strong G-band resonance was observed under 532 nm excitation and it was suppressed by carrier doping, resulting in an approximately threefold increase in $I_G/I_{2D}$ near the Dirac point [S7]. In 3D nanoporous graphene for Device 1, as shown in Fig. S9, lattice softening led to a reduction in the $I_G/I_{2D}$ ratio of only ~10%. In addition, only a small fraction of the entire sample area satisfies this resonance condition. Thus, the G-band intensity enhancement occurs only locally and appears as a weak intensity reduction near the Dirac point.

## 2. 4. G band softening of bilayer 2D graphene

In 3D-NPG EDLTs, when the number of graphene layers is three or higher, asymmetric charge doping occurs between the outermost layer and the interior layers (Figure 1c). A shift of $E_F$ relative to the Dirac point leads to G-band softening near the Dirac point, and a splitting of the G band is observed for Device 1. On the other hands, when the number of graphene layers is two in 3D-NPG EDLTs, the curvature radius of ~100 nm is sufficiently larger than the thickness of bilayer graphene. Thus, the amount of the charges is almost equally doped per unit area on the top and bottom layer and then no G-band splitting is expected. To verify this difference, a bilayer graphene-based EDLT was fabricated using a commercially available randomly stacked bilayer 2D graphene sheet with similar experimental conditions/materials of the 3D-NPG EDLTs. The graphene was sandwiched with gels (i.e. the top and bottom sides of the graphene was attached with the gels). The gate- and carrier-density dependences of the G-band spectra were measured under applied gate voltage $V_G$.

From the gate-voltage dependence on randomly stacked two-dimensional bilayer graphene EDLTs, around $V_G = -0.045$ V to 0.045 V, in which the electrical resistance shows a maximum, a minimum in the G-band frequency and a maximum in the full width at half maximum (FWHM) were observed. Upon both increasing and decreasing the $V_G$, a blue shift of the frequency and a decrease in FWHM were observed, but the peak splitting which was observed in 3D-NPG EDLT did not appear in the 2D graphene EDLTs. We derived $\omega_G$ and $\Gamma_G$ by using Lorentzian function fitting. These values were plotted against the carrier density estimated from capacitance measurements (Figure S13a, b). $\omega_G$ decreased and $\Gamma_G$ increased near the Dirac point owing to G-band softening. These results strongly suggested that the G-band splitting in 3D-NPG arises from asymmetric charge doping between the outermost and interior layers when the graphene has three or more layers.

## 2. 5. Estimations of carrier density from capacitance measurements

We determined the carrier density at each $V_G$ from capacitance measurements of the EDLT. For Device 1, with decreasing the frequency, the raw capacitance $C_s$ (i.e. recorded raw capacitance from Device 1 using impedance experiments) increased below 100 Hz and saturated, confirming the formation of the EDL capacitor. With further reductions in frequency, additional capacitance arose from impurity- and/or defect-related contributions (Figure S13a). In our analysis, we considered the $C_s$ value saturated at 1 Hz as the effective areal capacitance. Since this effective capacitance is the series combination of the channel electrode capacitance ($C_{ch}$, i.e. capacitance of 3D-NPG) and the gate-electrode capacitance ($C_G$), we calculated $C_{ch}$ by applying the sample's projected basal-area ratio. The resulting $V_G$ dependence of $C_{ch}$ for Device 1 showed a clear minimum at $V_G \sim 0.045$ V and this was closed to the $V_G = 0.135$ V, where $\omega_G$ showed the minimum (Figure S13b). We set $V_G = 0.135$ V as the Dirac point and determined the average carrier density per multilayer, $N$, by integrating $C_{ch}$ with respect to $V_G$ and dividing by the half of total internal surface area of the channels in 3D-NPG. Then, $N$ was divided by the average number of layers (2.8) to estimate the carrier density $n$.

## 2. 6. Estimations of electron-phonon coupling constant from the G-band analysis

We quantitatively evaluated the electron–phonon coupling strength between Dirac electrons and the G-mode phonon of the 3D-NPG for Device 1. Figures S14a,b showed the $\omega_G$ and $\Gamma_G$ as functions of the $E_F$. The carrier density $n$ of single layer was estimated from the carrier density of the multilayer divided by the average number of layers (2.8) and $E_F$ was calculated using $E_F = \hbar v \sqrt{\pi n}$ in the monolayer graphene. The $E_F$-dependent variations in $\omega_G$ and $\Gamma_G$ were described by the following equations [S6].

$$\omega_G = \omega_G^0 + \Delta\omega_G^A + \Delta\omega_G^{NA} \qquad (SE1)$$

$$\Delta\omega_G^{NA} = \int_{-\infty}^{\infty} \frac{[f(E - E_F) - f(E)]E^2 sgn(E)}{E^2 - \frac{(\hbar\omega_G^0)^2}{4}} dE \qquad (SE2)$$

$$\Gamma_G = \Gamma_G^0 + \frac{\lambda_\Gamma}{4}\omega_G^0\left[f\left(-\frac{\hbar\omega_G^0}{2} - E_F\right) - f\left(\frac{\hbar\omega_G^0}{2} - E_F\right)\right] \qquad (SE3)$$

In these expressions, $\omega_{G0}$ is the G-band frequency at $E_F = 0$, $\Delta\omega_G^A$ denotes the adiabatic shift due to the change in the lattice constant, and $\Delta\omega_G^{NA}$ and $\Delta\Gamma_G$ show the non-adiabatic frequency correction and broadening of the G-band due to the electron–phonon coupling, respectively. Here, $f(E)$ is the Fermi–Dirac distribution function, $\lambda_\Gamma$ is the electron–phonon coupling constant at the $\Gamma$ point, and $\Gamma_{G0}$ denotes the phonon linewidth arising from factors other than electron–phonon interaction. Since $\Delta\Gamma_G$ vanishes due to Pauli blocking when $| E_F |> (\hbar \omega_{G0})/2$, the model does not account for G-band peak broadening or splitting effects arising from inhomogeneous doping. Accordingly, the analysis was conducted in the gate-voltage range $V_G = -0.495$ V to $0.855$ V. No increases in $\Gamma_G$ from inhomogeneous carrier doping of the graphene multilayer was observed, while the electron–phonon–coupling–induced enhancement of $\Gamma_G$ was pronounced. In this carrier-density region, the variation of $\omega_G^A$ is negligibly small ($\sim$ $0.1$ cm$^{-1}$), and we therefore neglect the $\omega_G^A$ term in the fitting [S8]. In this analysis, we accounted for the G-band broadening arising from the $E_F$ distribution caused by electron–hole puddles by convolving the theoretical $\omega_G$ and $\Gamma_G$ curves with a Gaussian distribution of energy fluctuations $\delta E_F$ around the $E_F$. As a result, the fitting curves successfully reproduce the $E_F$ dependent $\omega_G$ and $\Gamma_G$ (shown as the red and blue lines, respectively), yielding $\delta E_F = 0.14 \pm 0.07$ eV ($\delta n \sim 1.2 \times 10^{12}$ cm$^{-2}$), $\lambda_\Gamma = 0.03 \pm 0.02$, $\Gamma_0 = 19 \pm 2$ cm$^{-1}$, $\omega_0 = 1578 \pm 5$ cm$^{-1}$.

## 2.7. Evaluation of charged-impurity scattering using electrical transport measurements

To evaluate charged-impurity scattering in 3D-NPG, we performed electrical-transport measurements. For a comprehensive study, we prepared two device types, Device 1 (as-prepared 3D-NPG with repeatably water washing process) and Device 2 (electrochemically etched using an EDLT at $V_G = -3$ V for 1 h at 240 K). Device 1 and Device 2 were fabricated from different growth batches but experimental conditions were similar. For Device 2, to clarify how electrochemical etching reduces charged impurities and affects transport, we also examined two unetched devices (Device 2-UE1 and Device 2-UE2). Device 1, Device 2, and Device 2-UE2 were fabricated using the ionic liquid and measured using a $^4$He cryostat equipped with an 18 T superconducting magnet (18T-SM, JASTEC), which has a 22 mm sample-stage diameter sufficient to mount the ionic liquid gated devices. For Device 2-UE2, we fabricated the device using the ionic liquid gel in order to perform EDLT measurements in the MPMS-XL5 (Quantum design), whose small 7 mm sample-stage diameter required ionic liquid gel preventing the leakage of the ionic liquid during measurements. As shown in Fig. S11, all devices exhibited ambipolar transport with a pronounced $R(V_G)$ peak and a sign reversal in the Hall resistance. The peak resistances were $\sim 100 - 120\ \Omega$ for Devices 2-UE1 and Device 2-UE2, $\sim 80\ \Omega$ for Device 1, and $\sim 75\ \Omega$ for Device 2, reflecting different impurity levels among the devices.

To probe electron-scattering by charged impurities in more detail, we set $V_G$ to the value giving the maximum resistance at 240 K (near the Dirac point) for each device and measured the magnetoresistance from 1.4 K to 240 K. For Device 1 (Fig. S15a), a positive magnetoresistance arose from the cyclotron motion of Dirac electrons on 3D tubular graphene at 200 K. Upon cooling, this positive magnetoresistance was gradually suppressed, and negative magnetoresistance due to weak localization dominated at 1.4 K [S9]. No quantum Hall effect was observed up to 15 T. For Device 2 (Fig. S15b), the low-temperature negative magnetoresistance was suppressed and the positive magnetoresistance was clearly recovered, indicating reduced charged-impurity scattering. To directly compare scattering strengths, we contrasted the low-temperature magnetoresistance among Device 1, Device 2, and Device 2-UE1 with same device fabrication processes and measurement conditions using the 18T-SM system. As shown in Fig. S15c, at 1.4 K the magnitude of the negative magnetoresistance decreased from 15 % (Device 2-UE1) to 12 % (Device 1) and 8 % (Device 2), consistent with suppressed electron scattering and a recovery of carrier mobility in Device 2. For Device 2 before etching (Device 2-UE2), we further examined the $V_G$ dependence of the magnetoresistance to capture signatures of charged-impurity scattering using the MPMS-XL5. As shown in Fig. S15d, Device 2-UE2 showed the negative magnetoresistance was maximized near the Dirac point at $V_G = -0.67$ V.

Magnetoresistance at 1.4 K for all devices was analyzed using a monolayer-graphene weak-localization model [S9], and the phase-relaxation field $B_\phi$, the intervalley-scattering field $B_i$, and the intravalley-scattering field $B_S$ are summarized in Fig. S16. The fits show that $B_s$ exceeds $B_i$ and $B_\phi$ by more than an order of magnitude at 1.4 K. $B_S$ decreases from ~ 3 T (Device 2-UE1) to ~ 2 T (Device 1) and further to ~ 1 T (Device 2), while in Device 2-UE2 it's peaks near the Dirac point. Because the charged-impurity scattering rate scales as $\sqrt{1/n}$ where the $n$ is carrier density [S10], it is strongest near the Dirac point, underscoring it's dominant contribution to $B_s$. Consequently, the observed magnetoresistance trends demonstrate that charged-impurity scattering is markedly reduced in Device 2 because of the electrochemical etching.

## 2.8. Gate voltage dependence of T-R curves

We show the T–R curves of Device 2-UE1 (before electrochemical etching) measured at different gate voltages in Fig. S18. In the electron-doped regime ($V_G$ =1.5 V) and the hole-doped regime ($V_G$= −1.5 V), the resistance curves become nearly temperature-independent (almost flat) above 10 K, where $R \propto \log T$ behavior due to weak localization is reduced. Furthermore, since the carrier densities corresponding to $V_G = 1.5$ V and $V_G = -1.5$ V are higher than the electron–hole puddle density, the changes in the resistance curves induced by applying these gate voltages can be regarded as independent of whether electrochemical etching is performed. On the other hand, for $V_G = -0.67$ V, while the resistance ratio of ~1.5 between 2 K and 150 K was higher than that in our previous report [S11] by tuning $E_F$ near the Dirac points by applications of $V_G$, representing the more insulating behavior, the T-R curve was still linear and Arrhenius-like curves (Fig. 5 in the main text) was masked by residual impurity-induced local doping. Therefore, in order to capture the intrinsic electrical transport near the Dirac points, we performed electrochemical etching to reduce the charged impurities on curved graphene channels as well as $E_F$ tuning by EDLT for electrical transport measurements.

## 2.9. Evaluation of graphene oxidation by Raman and XPS

We considered whether the Arrhenius-type $R(T)$ behavior arises from graphene oxidation. Table S1 lists the Raman $I_D/I_G$ ratios, the oxygen-to-carbon (O/C) atomic concentration ratios from XPS, and the transport-derived excitation-gap values ($\Delta$) for Device 2. For the $I_D/I_G$ intensity ratio, which reflects the density of oxidation-induced sp3-like defects, Device 2 show $I_D/I_G = 0.13$, lower than zero-gap CVD graphene ($I_D/I_G = 0.19$) and far below oxidized CVD graphene with a 0.11–0.13 eV gap ($I_D/I_G = 0.61 - 0.83$) [S12]. By contrast, based on the fractional intensity of oxygen-containing functional groups in the C 1s spectra, the O/C ratios of Device 2 is 0.09, slightly higher than that of zero-gap CVD graphene (0.03) and approaching graphene-oxide levels (0.13–0.15) [S12]. This apparent contradiction between the $I_D/I_G$ and O/C ratios suggests additional oxidation of 3D-NPG during electron-beam irradiation during the XPS measurements.

To clarify this, we prepared three device conditions from the same batch of Device 2, as-grown device (before ionic-liquid immersion), IL-immersion device (after immersion in an ionic liquid), and EC-etching device (after electrochemical etching), and compared their C 1s spectra. As shown in Fig. S19 a, the fractional intensity of oxygen-containing functional groups in the C 1s spectra of the as-grown device was 11% (O/C = 0.04), consistent with zero-gap CVD graphene. It increased to ~34% in the IL-immersed device and ~29% in the EC-etched device. The higher oxygen intensity in the IL-immersed device than that in the EC-etched device indicated that the observed oxidation peaks arise not from electrochemical oxidation but from electron-beam–induced secondary oxidation of residual ionic liquid and surface amorphous carbon generated during the XPS measurements. Indeed, the S 2p spectra showed a clear peak in the IL-immersed device, attributable to residual DEME-TFSI on the 3D-NPG surface. Furthermore, the low $I_D/I_G$ ratio of 0.12–0.13 observed in these three devices in Fig. S19b supports this scenario. Thus, we conclude that the slightly elevated O/C ratio originates from oxidation of the ionic liquid and/or amorphous carbon on the 3D-NPG surface and does not account for the Arrhenius-type $R(T)$ behavior.

## 2. 10. Effects of finite-size quantization associated with the tube diameter and the tube length

For carbon nanotubes, finite-size quantization associated with the tube diameter and the tube length has been discussed extensively, and the corresponding characteristic energy scales can be estimated as the equation (diameter quantization [S15])

$$E_d = \frac{2t_0 r_0}{d}$$

and the equation (length quantization [S16])

$$E_L = \frac{\hbar v_F}{2L}.$$

Here, $t_0$ is the nearest-neighbor hopping integral, $r_0$ is the C–C bond length, $d$ is the tube diameter, $v_F$ is the Fermi velocity, and $L$ is the tube length. Using representative graphene parameters $t_0 = 3$ eV, $r_0 = 0.142$ nm, and $v_F = 1 \times 10^6$ m/s, together with the characteristic geometrical scales inferred from our STEM images ($d \sim 100 - 150$ nm and $L \sim 200 - 400$ nm), we obtain $E_d \sim 3 - 4$ meV and $E_L \sim 1 - 2$ meV. These energy scales are about one order of magnitude smaller than the gap value estimated from Eq. (1) in the main text (48 meV), indicating that confinement-related quantization is unlikely to be the dominant contribution to the observed transport gap.

Because the tube diameter varies from place to place in our 3D-NPG network, the confinement energy arising from the diameter is spatially distributed, which in turn reduces the overall electronic confinement effect. Moreover, unlike an isolated nanotube with well-defined boundary conditions, 3D-NPG forms a monolithic and bicontinuous network like spider nets, so the electronic confinement effect is further weakened. Therefore, the length quantization should be even more broadened and difficult to be observed experimentally.

## 2. 11. Effects of domain boundary for the electrical transport

From our HR-TEM observations, we have not been able to clearly observed well-defined domain boundaries as reported in Ref. S17. Considering that the domain boundaries could be formed by adjustments of graphene lattice mismatches by geometric requests, our nanoporous graphene could release the geometric frustration by forming a lot of single or pair topological defects on the curved regions. This could be attributed to avoid the formation of such large domain boundaries.

Moreover, our transport results suggest that domain-boundary scattering is unlikely to be the dominant factor assuming the electron mean free path. If domain boundaries were the primary limitation, one would not expect a substantial change in (i) the absolute magnitude of the conductivity and (ii) the qualitative magnetoresistance (MR) behavior before and after electrochemical etching. Experimentally, however, after etching we observe a decrease in resistance together with a more clearly band-like MR response characteristic of graphene, indicating that charged-impurity-related scattering is the dominant contribution to electron scattering in our devices.

Based on these discussions, since (i) domain boundaries are not clearly resolved in our HR-TEM images, (ii) realistic boundaries are expected to have structural fluctuations and finite segment lengths that relax the strict momentum-selection rules, (iii) domain-boundary scattering is unlikely to be the dominant for electrical transport in our devices, these provides minor contributions. At minimum, the absence of long, highly symmetric, and well-ordered boundaries in our observations suggests that transport-gap formation dominated by domain boundaries is not the main mechanism in the 3D-NPG.

## 2. 12. Estimation of density of topological defects from associated with spherical (positive) curvature

We provide an order-of-magnitude estimate for the density of topological defects associated with spherical (positive) curvature, following the approach discussed in Ref. S18.

For a curved graphene patch *D*, Gauss–Bonnet relates the integrated Gaussian curvature to a topological invariant as

$$\iint_D K \, dA + \oint \kappa_g ds = 2\pi.$$

To connect curvature to non-hexagonal rings, we use the standard correspondence that a pentagon (removal of a 60° wedge) introduces a positive curvature of $+\frac{\pi}{3}$, while a heptagon (insertion of a 60° wedge) introduces a negative curvature of $-\pi/3$. Then the net topological "charge" satisfies

$$\frac{\pi}{3}(N_5 - N_7) = \iint_D K \, dA,$$

where $N_5$ and $N_7$ are the numbers of pentagons and heptagons within the patch, respectively.

As a representative example, we consider a spherical cap of radius $R = 100$ nm with opening angle $\theta = 10°$. For a sphere, $K = \frac{1}{R^2}$, and thus

$$\iint_D K \, dA = \int_0^{2\pi} \int_0^{\theta} \frac{1}{R^2} R^2 \sin\theta \, d\theta d\phi = 2\pi(1 - \cos\theta).$$

With $\theta = 10°$, this gives $\iint_D K \, dA \sim 0.095$ rad, and therefore $N_5 \sim 0.09$. Assuming that this positively curved patch is mainly accommodated by pentagons (i.e., $N_7 \sim 0$ locally), we obtain minimum $N_5 \sim 0.09$ within this area, i.e., substantially less than one pentagon per such patch on average. The area of the same spherical cap is

$$\int_0^{2\pi} \int_0^{\theta} R^2 \sin\theta \, d\theta d\phi = 2\pi R^2 (1 - \cos\theta) \sim 9.55 \times 10^2 \text{nm}^2.$$

Using the graphene unit-cell area $A_{uc} = 0.0524$ nm$^2$ (two carbon atoms per unit cell), the number of carbon atoms in this patch is approximately $N = 2 \times \frac{A}{A_{uc}} \sim 3.65 \times 10^4$. Therefore, the corresponding pentagon fraction is of order

$$\frac{N_5}{N_{atom}} \sim \frac{0.09}{3.65 \times 10^4} \sim 2.5 \times 10^6,$$

i.e., $\sim 3 \times 10^{-4}$ %. This is far below a representative high defect density such as $\sim 0.1$ %. This order-of-magnitude estimate suggests that the contribution of curvature-required topological defects is expected to be highly dilute, and thus the gapping-out of Dirac fermions associated with such defects is more naturally interpreted as a local effect in limited regions of the 3D labyrinth rather than a globally averaged effect.

We also note an important limitation of curvature-based counting. Defects such as the 5–8–5 structure also highlight an important limitation of curvature-based counting. In the standard wedge-disclination picture, a pentagon contributes a positive curvature of $+\frac{\pi}{3}$, whereas an octagon contributes a negative curvature of $-\frac{2\pi}{3}$. Therefore, a 5–8–5 defect carries a net topological curvature charge of $\frac{\pi}{3} - \frac{2\pi}{3} + \frac{\pi}{3} = 0$. In such cases, estimating defect density from the average or integrated curvature (via Gauss–Bonnet-type arguments) becomes intrinsically insensitive and may systematically underestimate the presence of these defects. Considering (i) the order-of-magnitude curvature-based estimate above, (ii) the limited number of directly identified non-hexagonal rings in our STEM images, and (iii) the coexistence of an insulating channel and itinerant Dirac-electron channel in transport, we conclude that gap formation associated with topological defects is most consistently interpreted as a predominantly local effect rather than a globally uniform one.

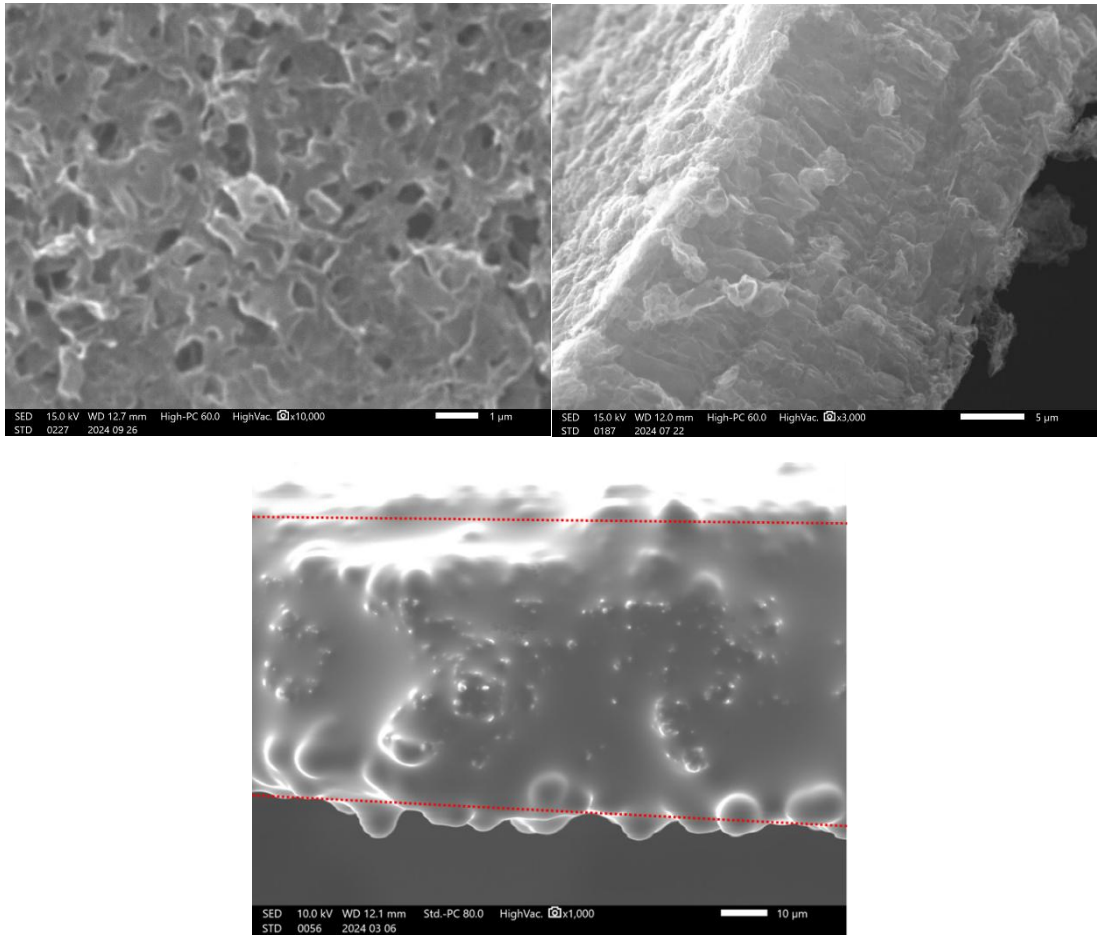

**Figure S1**. SEM images of 3D nanoporous graphene before (top) and after (bottom) filling the gel. Red dot lines presented the edge of graphene sheet.

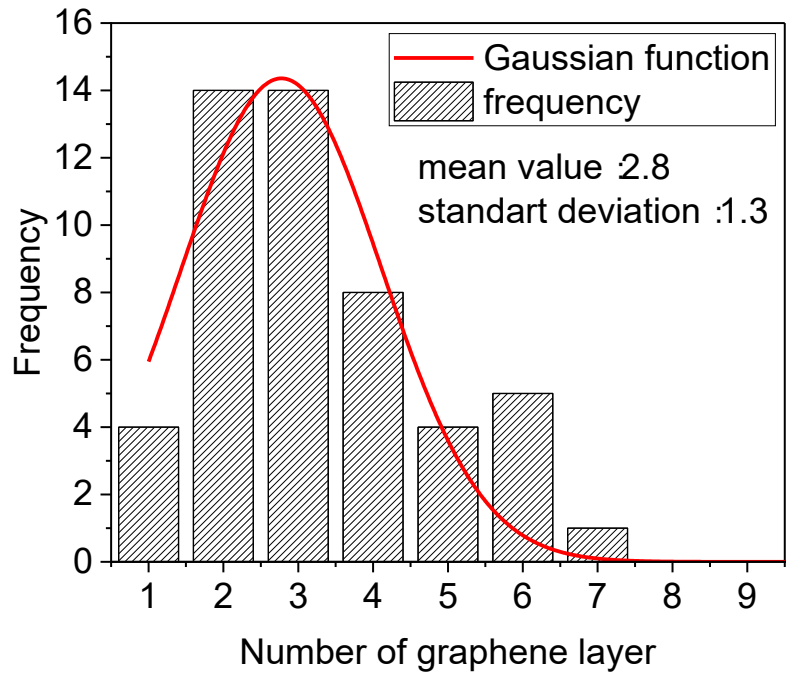

**Figure S2.** Analysis of the number of graphene layers in the 3D nanoporous graphyene. The number of layers was counted on the edges from several TEM images, and a Gaussian function was fitted to estimate the layer number as approximately 2 – 4.

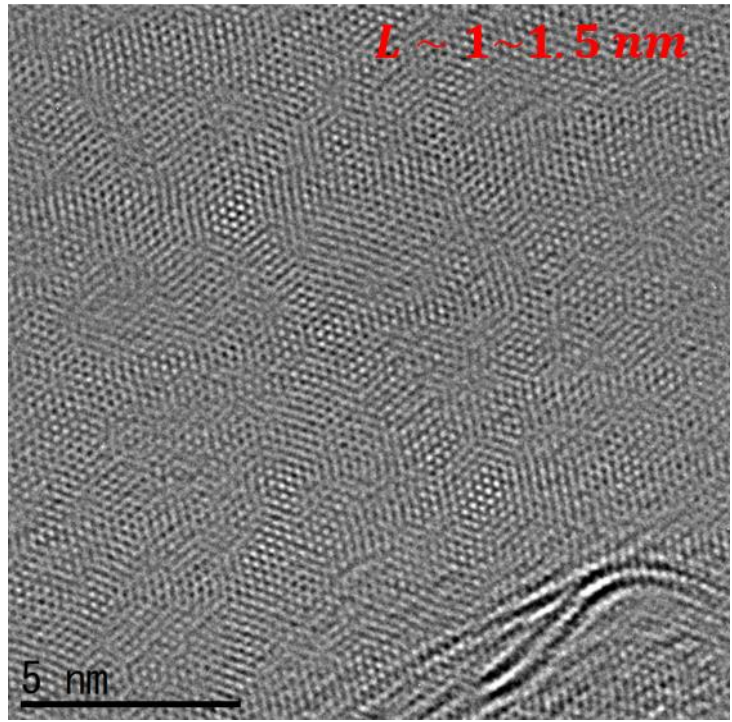

**Figure S3.** HR-TEM image of 3D nanoporous graphene. A modulation of the moiré length scale was observed in the range $L = {\sim}1 - 1.5$ nm and $\theta \sim 9° - 14°$. Moreover, beyond the primary moiré periodicity, long-range motifs became apparent. Sharply defined AA-stacked regions alternated with more diffuse AB-stacked domains, arising from interference among randomly stacked multilayers ("a moiré of moirés").

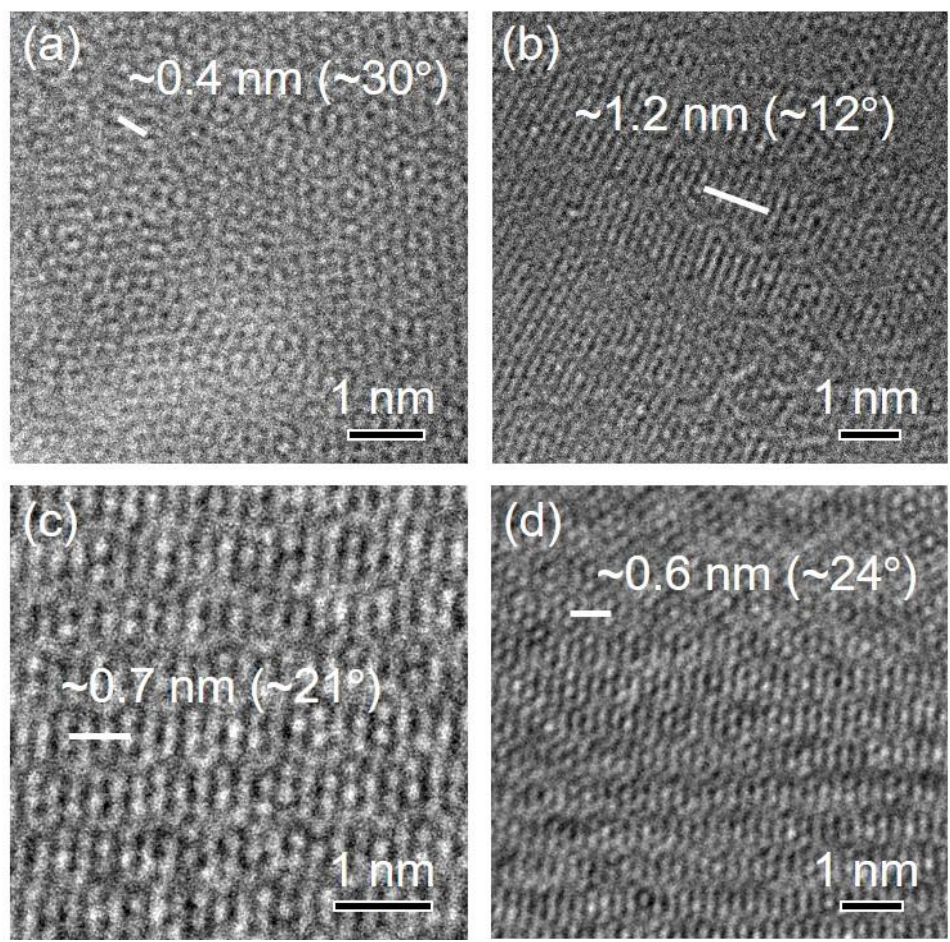

**Figure S4.** High-resolution TEM images of 3D nanoporous graphene with curvature radius of (a, b) 25-50nm and (c,d) 500-1000 nm.

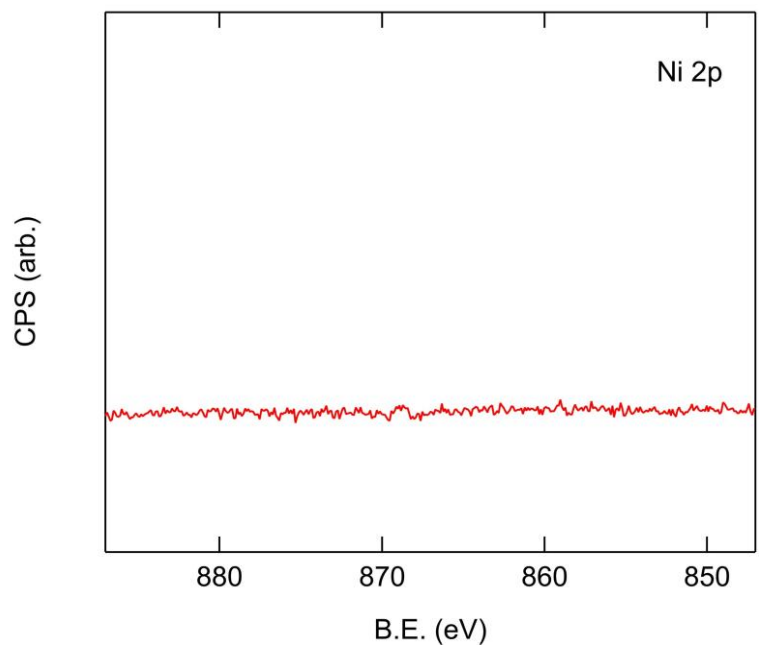

**Figure S5**. XPS spectra of Ni 2p on 3D nanoporous graphene. The residual Ni atomic concentration was not obviously detected.

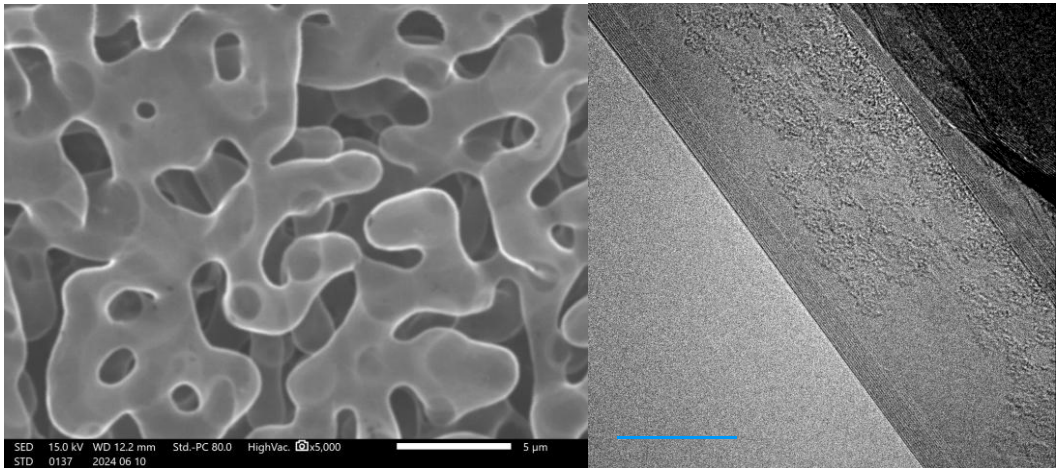

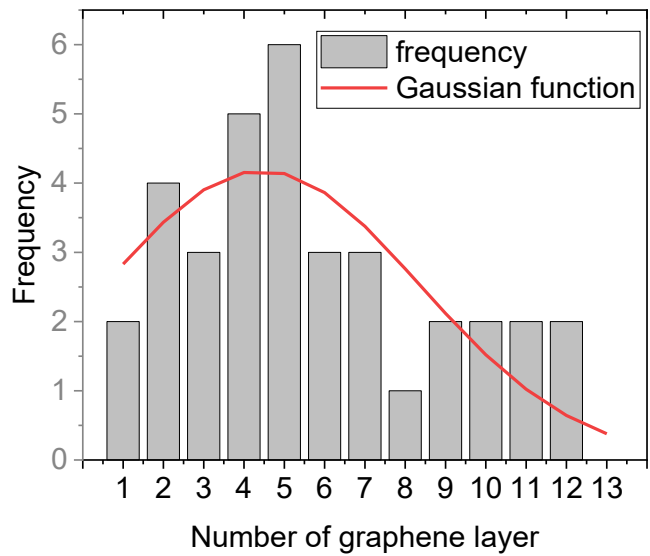

Figure S6. Morphology of NPG with 500-1000 nm radius and the layer distribution analyzed from several TEM images (G950).

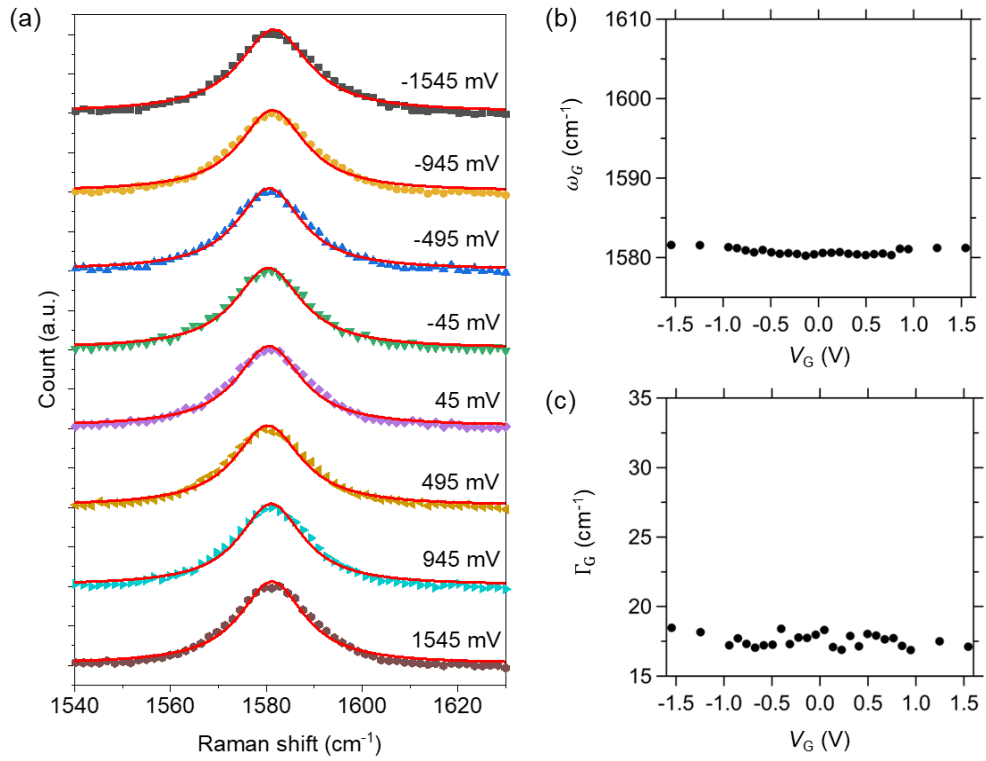

**Figure S7.** Gate voltage ($V_G$) dependence of Raman G-band of G950 with curvature radius of 500-1000 nm. (a) G-band spectra at various $V_G$. The Raman spectra were normalized for easy comparison. (b) $V_G$ dependence of G-band frequency ($\omega_G$). (c) $V_G$ dependence of G-band line width ($\Gamma_G$).

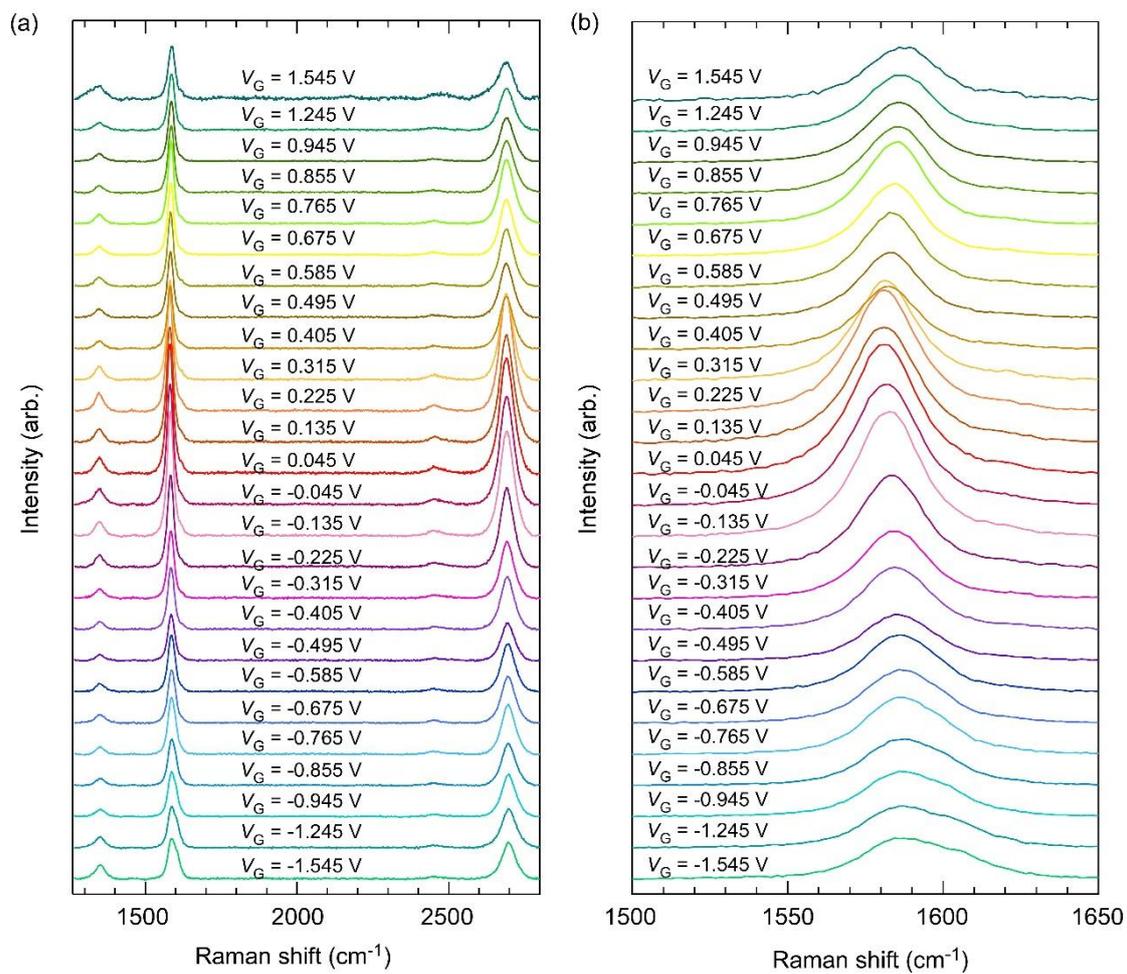

**Figure S8**. Gate voltage ($V_G$) dependence of Raman G-band of G800 with curvature radius of 100-150 nm. (a) Raman spectra of 3D nanoporous graphene EDLT in the gate voltage range from −1.545 V to 1.545 V using Device 1. (b) Magnification of G-band spectra in the gate voltages. The Raman spectra were not normalized.

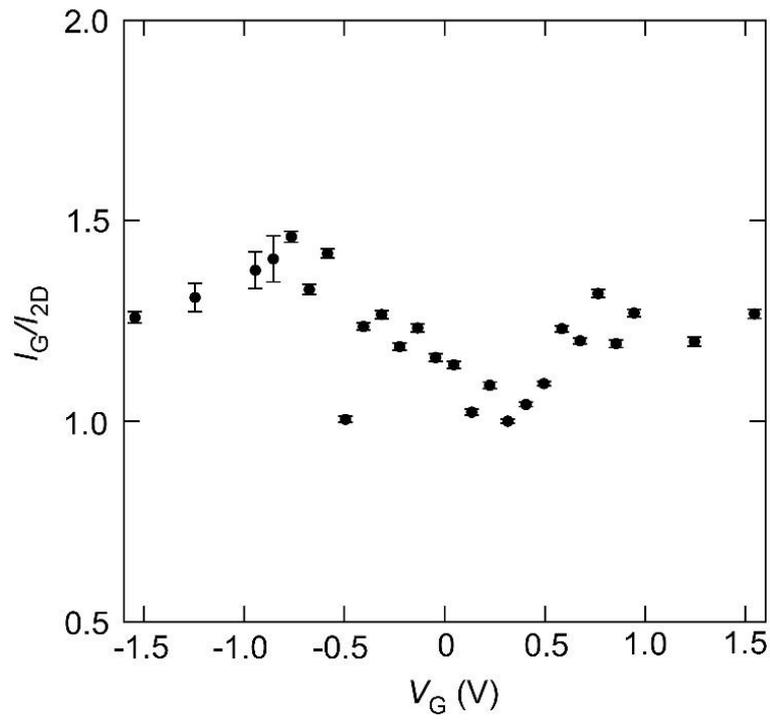

**Figure S9.** The intensity ratio of G band and 2D band ($I_G/I_{2D}$) in the gate voltage ($V_G$) from $-1.545$ V to $1.545$ V using Device 1 (G800 with curvature radius of 100-150 nm).

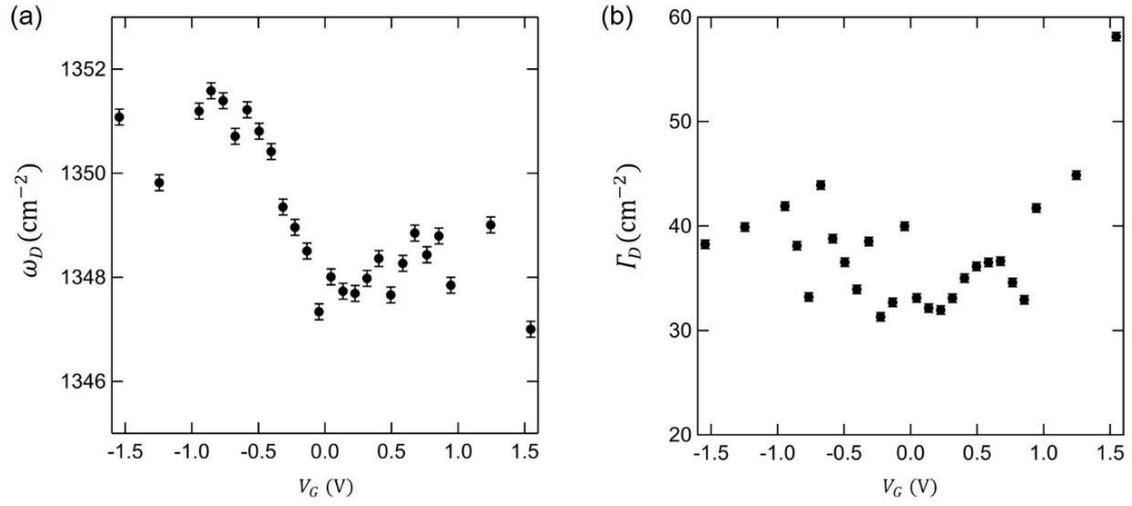

**Figure S10**. D-band characters in Raman spectra in the gate voltage ($V_{\mathrm{G}}$) from $-1.545$ V to $1.545$ V using Device 1 (G800 with curvature radius of 100-150 nm). (a) D-band peak position $\omega_D$, (b) D-band full width at half maximum $\Gamma_D$. $\omega_D$ and $\Gamma_D$ were estimated from the spectra fitting using the Lorentzian function. $V_G$ dependence of $\omega_D$ and $\Gamma_D$ was resemble to the theoretical calculations of chemical potential dependence of real and imaginary part of self-energy for the TO optical phonon with $vq = 1$ eV [S13].

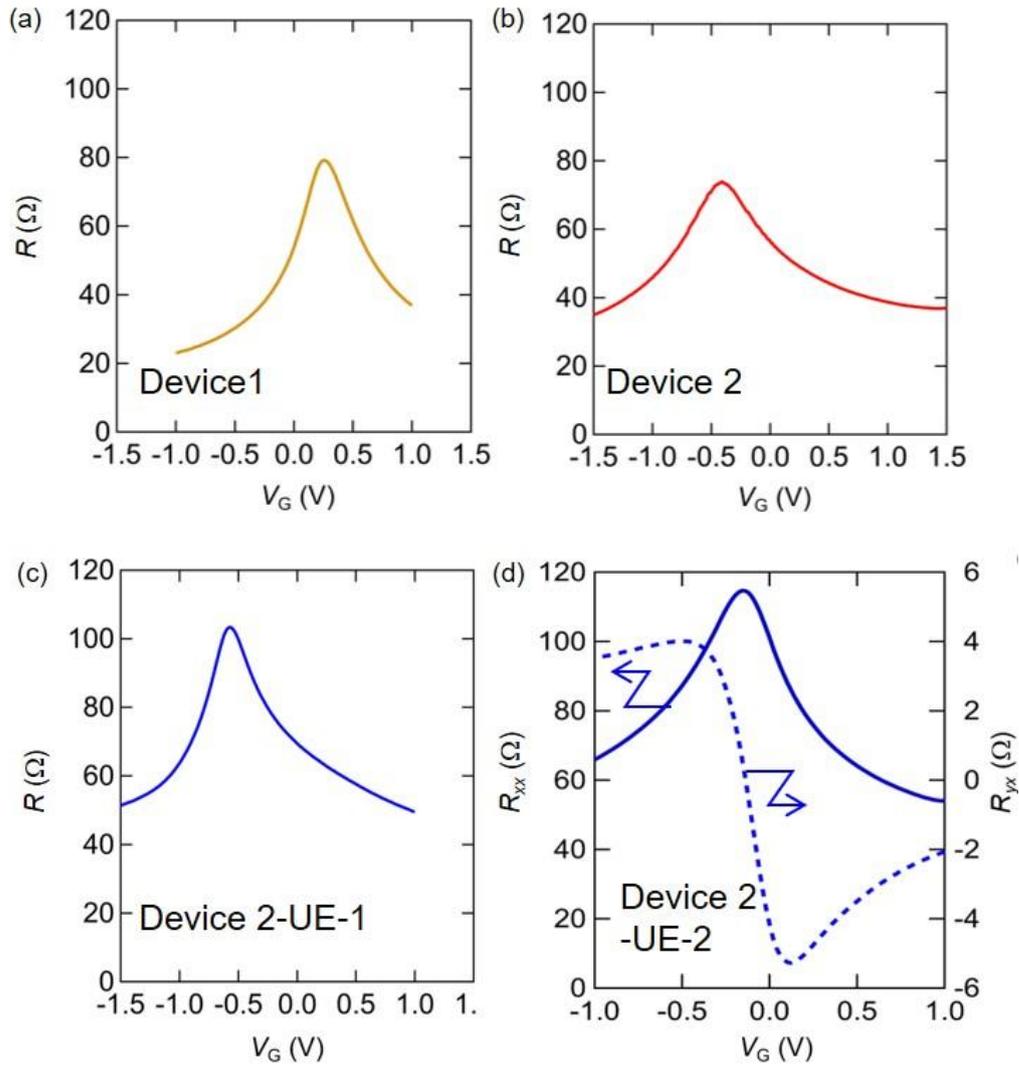

**Figure S11.** Gate voltage ($V_G$) dependence of longitudinal and Hall resistance for 3D nanoporous graphene EDLT. (a) Device 1, (b) Device 2, and (c, d) Device 2 before electrochemical etching ((c) : Device 2-UE-1, (d) : Device 2-UE-2). In (d), the solid line is $R_{xx}$ at $B = 0$ T and the broken line is $R_{yx}$ at $B = 15$ T. Device 1 and Device 2 were fabricated using different growth batches. Device 2 in (b-d) different devices using same 3D-NPG growth batch.

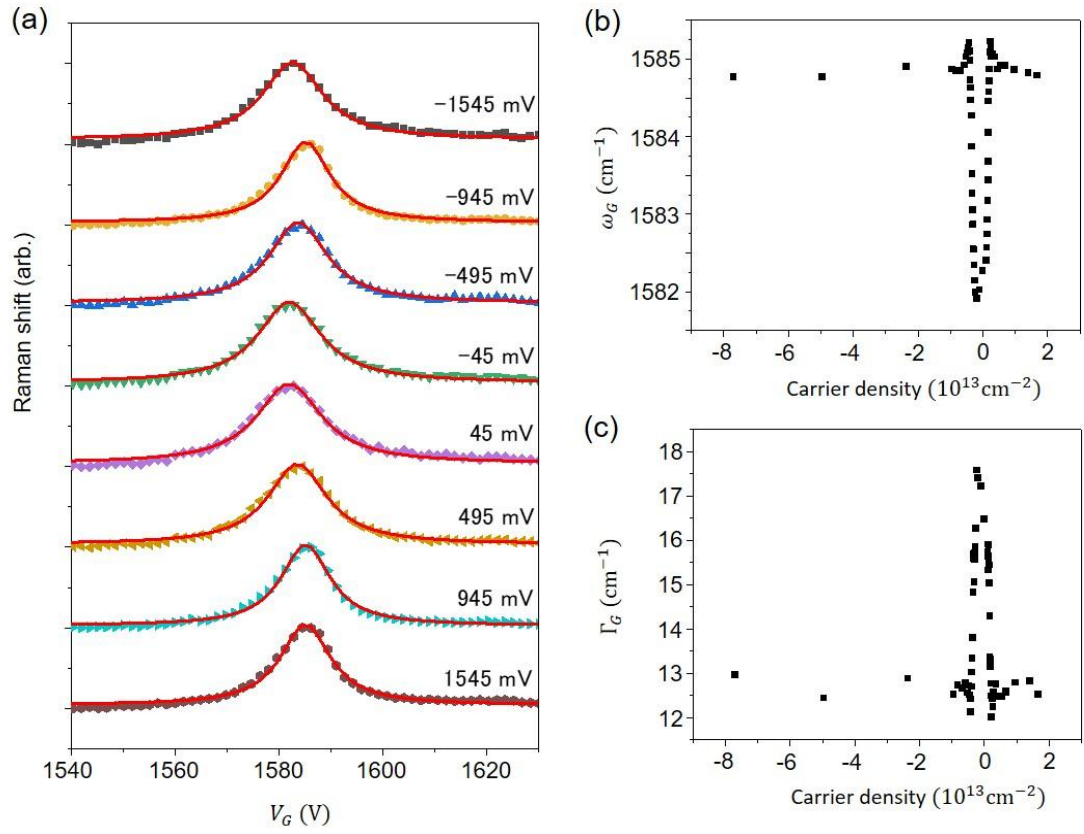

**Figure S12**. Raman spectra of randomly two-dimensional twisted bilayer graphene EDLT (2D bilayer graphene EDLT). (a) Raman spectra in the gate voltage range from −1.545 V to 1.545 V. Solid lines were fitting curves using the Lorentzian function. The Raman spectra were normalized for easy comparison. $V_G$ dependence of (b) G-band frequency, $\omega_G$, and (c) full width at half maximum $\Gamma_G$.

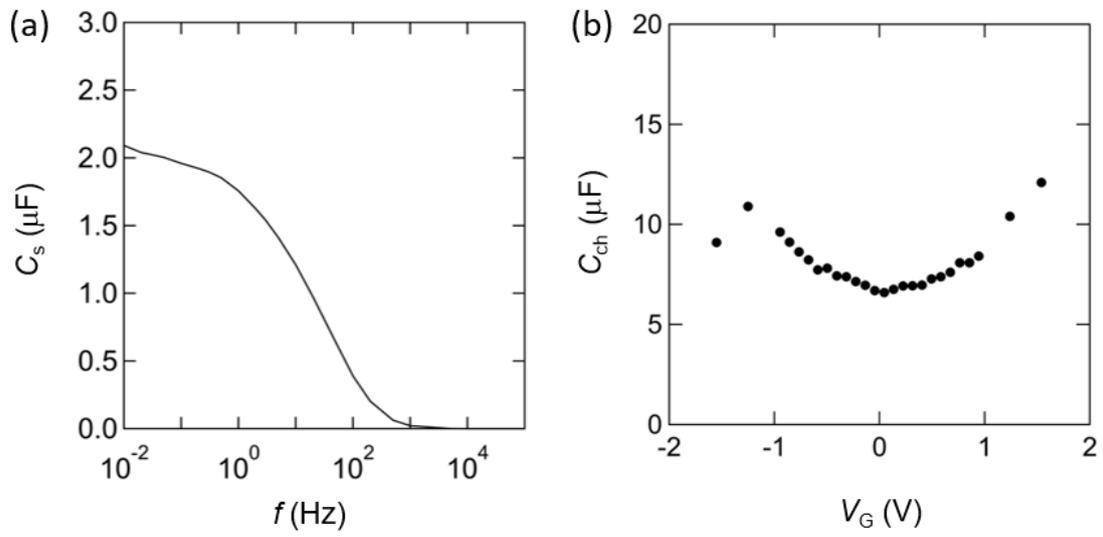

**Figure S13.** (a) Typical frequency dependence of raw capacitance $C_s$ estimating from impedance measurements of 3D nanoporous graphene EDLT using Device 1. (b) Gate-voltage ($V_G$) dependence of the channel electrode capacitance ($C_{ch}$) of Device 1.

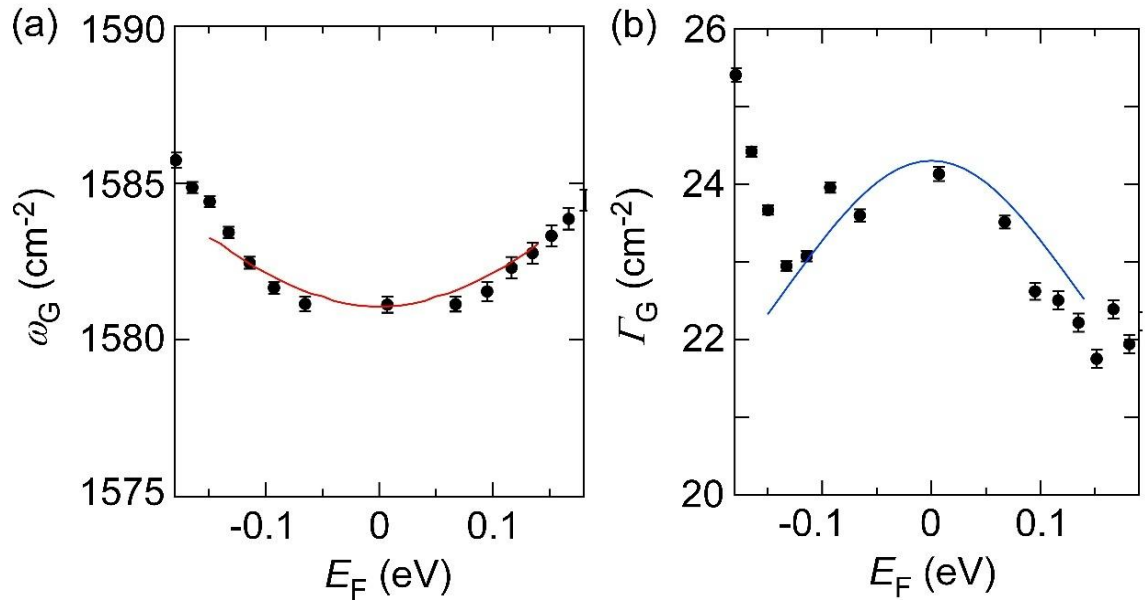

**Figure S14.** The averaged $E_F$ dependence of (a) $\omega_G$ and (b) G-band FWHM $\Gamma_G$. $E_F$ was calculated using $V_G$ dependence of EDL capacitance with the average of graphene layer (2.8) using Device 1. Red and blue lines were fitting results of equation (SE1-3).

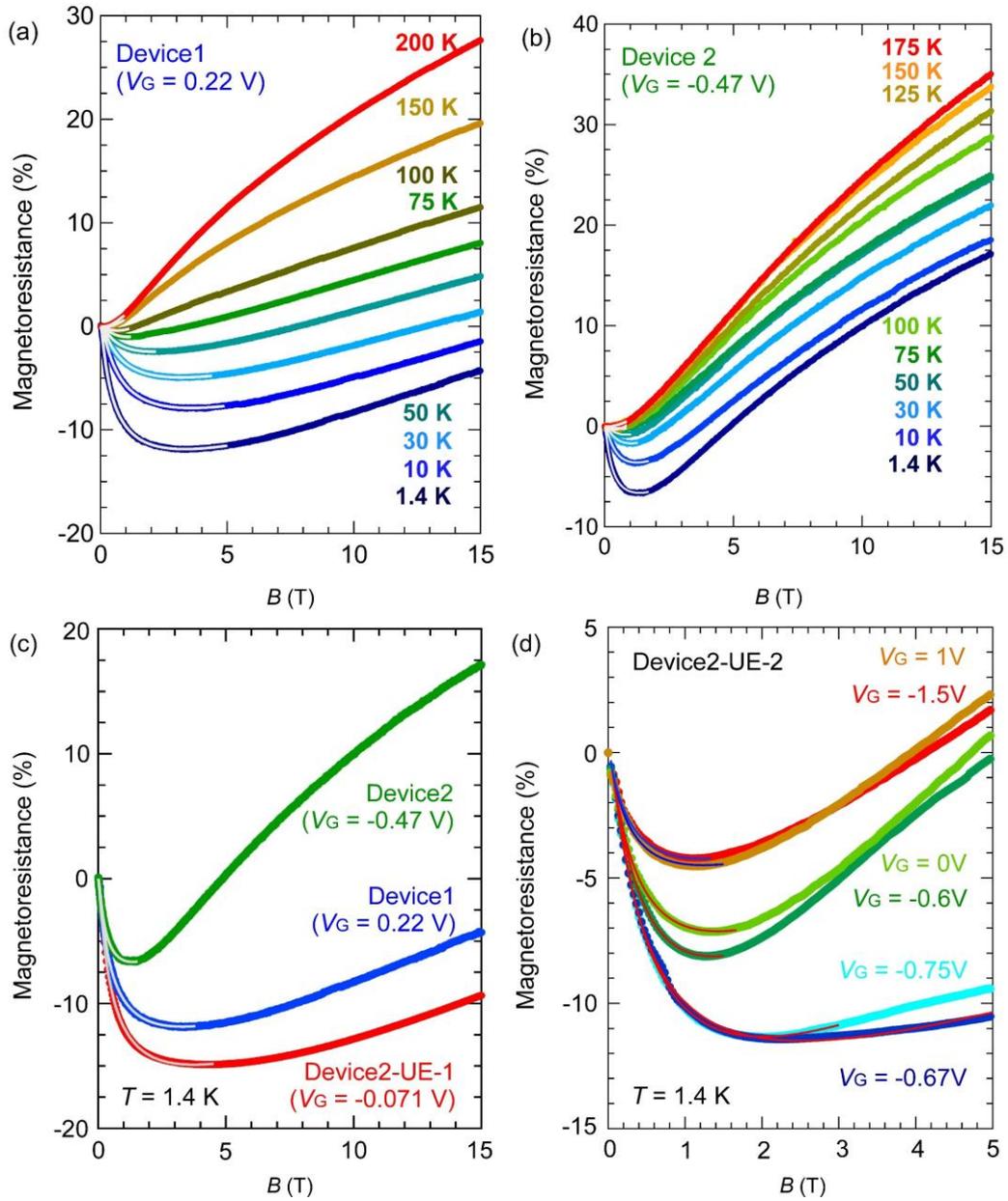

**Figure S15.** Magnetoresistance of 3D nanoporous graphene EDLTs of (a) Device 1 and (b) Device 2 in the temperature ranges from 1.4 K to 200 K near the Dirac points. (c) Magnetoresistance of 3D-NPG-EDLTs (Device 1, Device 2, Device 2-UE-1) measured at 1.4 K near the Dirac points. (d) Gate-voltage dependence of the magnetoresistance of Device 2-UE-2 at 1.4 K. $V_G = -0.67$ V near the Dirac point. The solid lines in (a-d) represented fitting results using a monolayer graphene weak-localization model [S9].

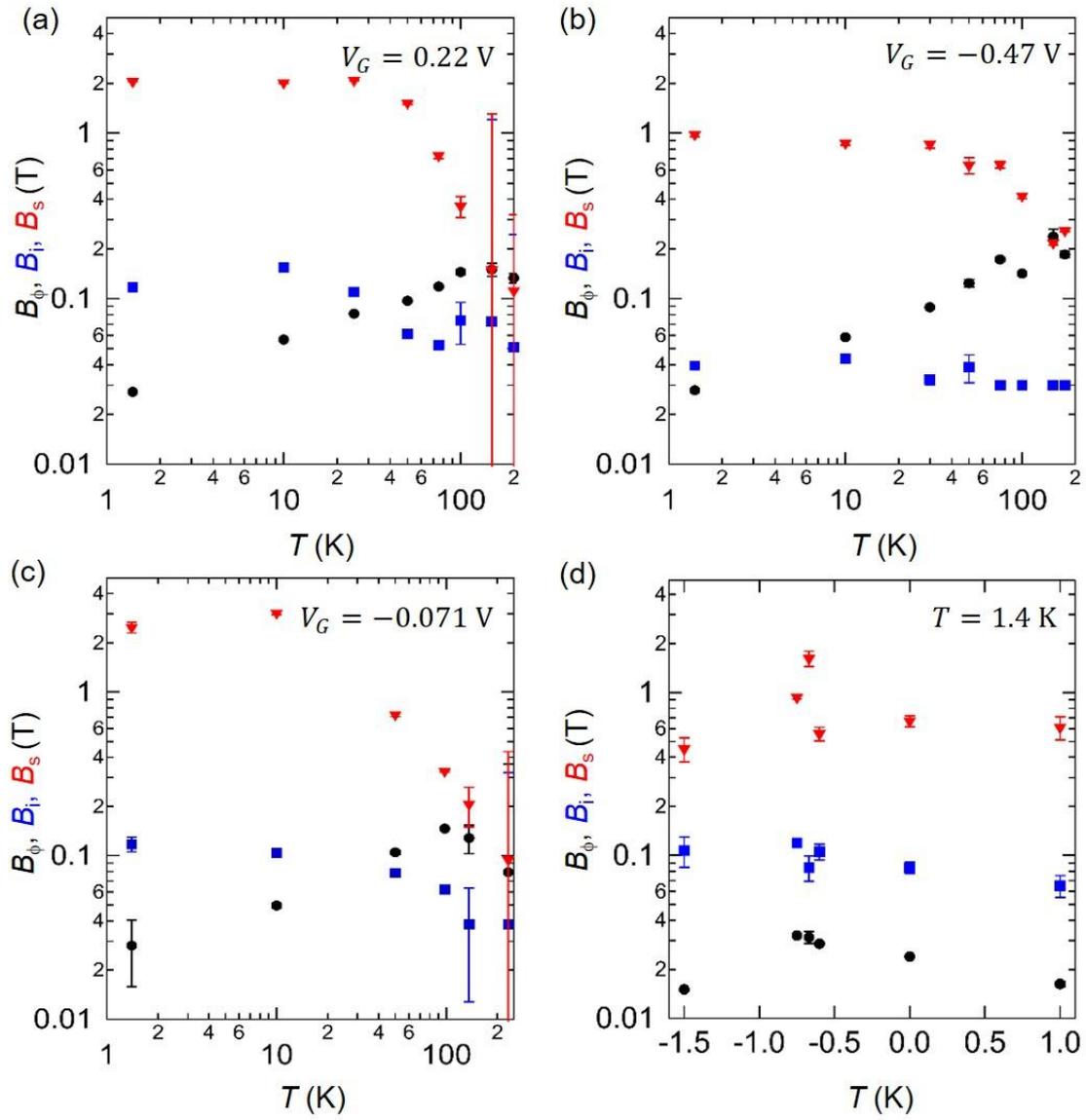

**Figure S16.** Analyses of the negative magnetoresistance from 1.4 K to 240 K near the Dirac points using a monolayer graphene weak-localization model [S9]. (a) Device 1, (b) Device 2, (c) Device 2-UE-1. $B_\phi$ (black circles), $B_i$ (blue squares), and $B_s$ (red triangles) denoted the phase relaxation field, intervalley scattering field, and intravalley scattering field, respectively. (d) $V_G$ dependence of $B_\phi$, $B_i$, and $B_s$ at 1.4 K for Device 2-UE-2. $V_G = -0.67$ V near the Dirac points.

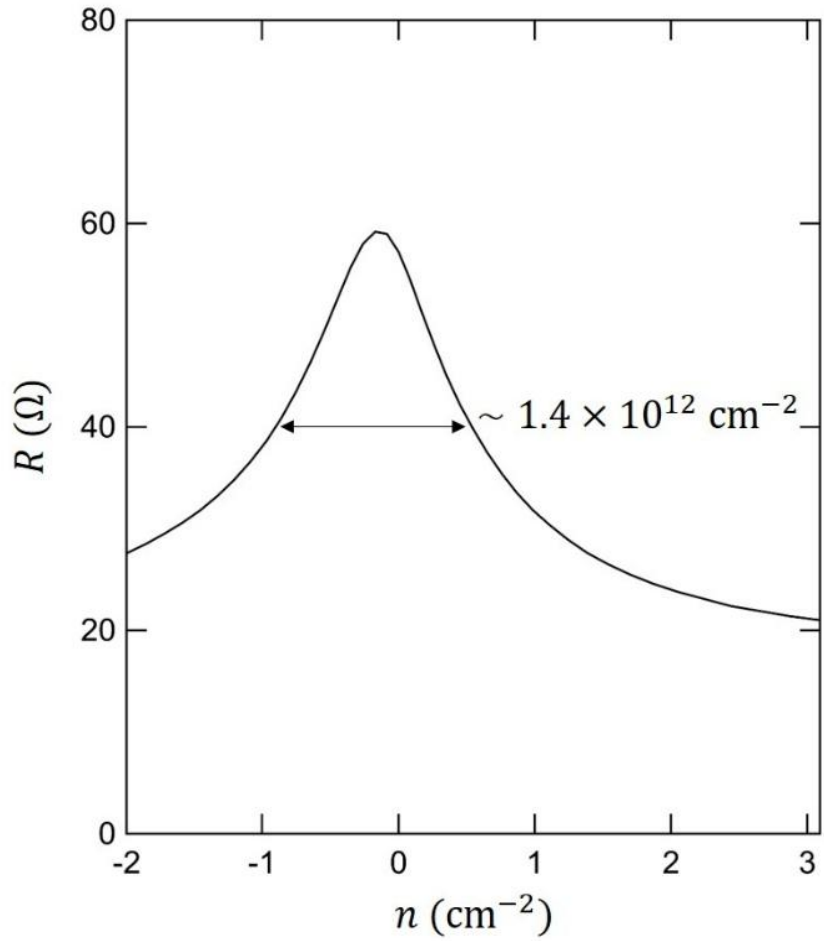

**Figure S17.** *R* as a function of *n* for Device 2 (This device is a different device from the same batch as Device 2 in the main text.) The carrier density was obtained from the EDL capacitance by integrating over the $V_G$ and normalizing by one half of the total internal channel surface area and by the average number of layers (2.8). The electron–hole puddle density, $\delta n$, was estimated from the full width at half maximum (FWHM) of the resistance peak [S14].

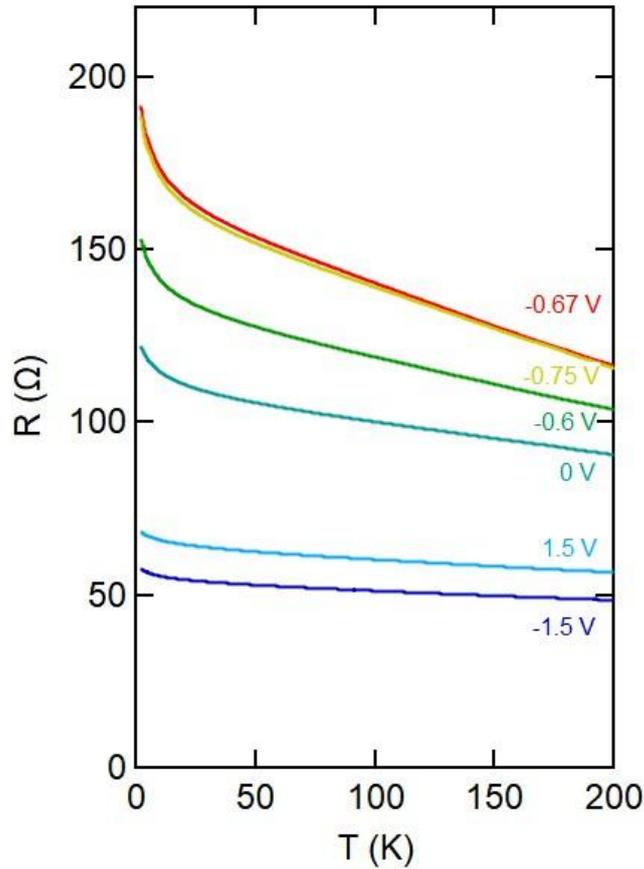

**Figure S18.** Temperature-Resistance (T–R) curves for Device2-UE1 under various gate voltages. Device2-UE1 denotes an Device2 without electrochemical etching process but with tuning of Fermi level. Since the carrier densities corresponding to $V_G$ = 1.5 V and $V_G = -1.5$ V are much higher than the electron–hole puddle density, the behaviors in the resistance curves induced by applying these gate voltages between Device 2 before and after electrochemical etching could be almost identical.

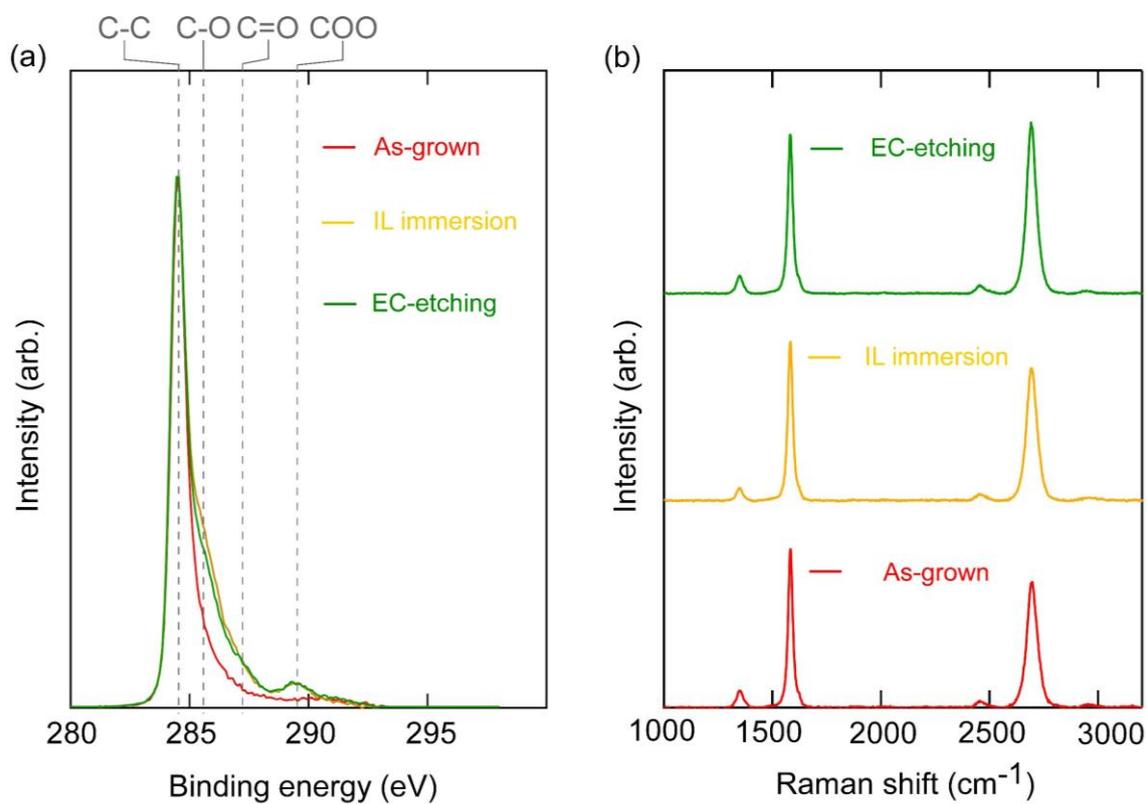

**Figure S19.** (a) XPS and (b) Raman spectra of Device 2. Different state of Device 2 such as as-grown state (before ionic-liquid immersion), IL-immersion state (after immersion in an ionic liquid), and EC-etching state (after electrochemical etching) were compared.

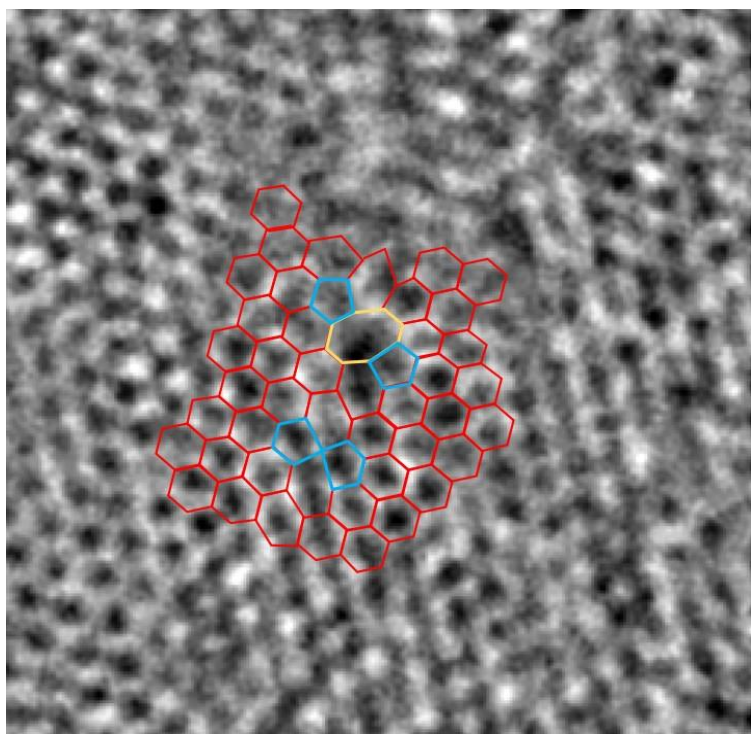

**Figure S20.** Typical high-resolution TEM image of 3D nanoporous graphene around topological defects at a curved region. Five- and eight-membered ring structures were observed.

TableS1. Summary of the $I_D/I_G$ ratios from Raman spectroscopy, oxygen-to-carbon atomic concentration ratios (O/C) from XPS, and band gaps from electrical-transport measurements for Device 2. For comparison, CVD graphene and graphene oxide data were also summarized[S12].

|  | Device2 | CVD- graphene [S12] | Graphene-oxide [S12] |
|---|---|---|---|
| $I_D/I_G$ | 0.12 | 0.19 | $0.61 - 0.83$ |
| $O/C$ | 0.09 | 0.03 | 0.13 |
| Band Gap (eV) | 0.05 | 0 | 0.11 |

# References

[S1] Y. Tanabe, Y. Ito, K. Sugawara, S. Jeong, T. Ohto, T. Nishiuchi, N. Kawada, S. Kimura, C.F. Aleman, T. Takahashi, M. Kotani, M. Chen, Coexistence of Urbach-tail-like localized states and metallic conduction channels in nitrogen-doped 3D curved graphene, Adv. Mater. 34 (2022) 2205986. https://doi.org/10.1002/adma.202205986

[S2] N. Nakatsuji, T. Kawakami, M. Koshino, Multiscale lattice relaxation in general twisted trilayer graphenes, Phys. Rev. X 13 (2023) 041007. https://doi.org/10.1103/PhysRevX.13.041007

[S3] P. Moon, M. Koshino, Optical absorption in twisted bilayer graphene, Phys. Rev. B 87 (2013) 205404. https://doi.org/10.1103/PhysRevB.87.205404

[S4] Z. Zhu, S. Carr, D. Massatt, M. Luskin, E. Kaxiras, Twisted trilayer graphene: A precisely tunable platform for correlated electrons, Phys. Rev. Lett. 125 (2020) 116404. https://doi.org/10.1103/PhysRevLett.125.116404

[S5] Y. Ito, Y. Tanabe, H.-J. Qiu, K. Sugawara, S. Heguri, N.H. Tu, K.K. Huynh, T. Fujita, T. Takahashi, K. Tanigaki, M. Chen, High-quality three-dimensional nanoporous graphene, Angew. Chem. Int. Ed. 53 (2014) 4822–4826. https://doi.org/10.1002/anie.201402662

[S6] G. Froehlicher, S. Berciaud, Raman spectroscopy of electrochemically gated graphene transistors: Geometrical capacitance, electron–phonon, electron–electron, and electron–defect scattering, Phys. Rev. B 91 (2015) 205413. https://doi.org/10.1103/PhysRevB.91.205413

[S7] T.F. Chung, R. He, T.L. Wu, Y.P. Chen, Optical phonons in twisted bilayer graphene with gate induced asymmetric doping, Nano Lett. 15 (2015) 1203–1210. https://doi.org/10.1021/nl504318a

[S8] M. Lazzeri, F. Mauri, Nonadiabatic Kohn anomaly in a doped graphene monolayer, Phys. Rev. Lett. 97 (2006) 266407. https://doi.org/10.1103/PhysRevLett.97.266407

[S9] E. McCann, K. Kechedzhi, V.I. Fal'ko, H. Suzuura, T. Ando, B.L. Altshuler, Weak-localization magnetoresistance and valley symmetry in graphene, Phys. Rev. Lett. 97 (2006) 146805.

[S10] E.H. Hwang, S. Adam, S. Das Sarma, Carrier transport in two-dimensional graphene layers, Phys. Rev. Lett. 98 (2007) 186806. https://doi.org/10.1103/PhysRevLett.98.186806

[S11] Y. Tanabe, Y. Ito, K. Sugawara, M. Koshino, S. Kimura, T. Naito, I. Johnson, T. Takahashi, M. Chen, Dirac Fermion Kinetics in 3D Curved Graphene, Adv. Mater. 32 (2020) 2005838. https://doi.org/10.1002/adma.202005838

[S12] A.I. Aria, A.W. Gani, M. Gharib, Effect of dry oxidation on the energy gap and chemical composition of CVD graphene on nickel, Appl. Surf. Sci. 293 (2014) 1–11. https://doi.org/10.1016/j.apsusc.2013.11.117

[S13] K. Sasaki, K. Kato, Y. Tokura, S. Suzuki, T. Sogawa, Decay and frequency shift of both intervalley and intravalley phonons in graphene: Dirac-cone migration, Phys. Rev. B 86 (2012) 201403(R). https://doi.org/10.1103/PhysRevB.86.201403

[S14] Y.-W. Tan, Y. Zhang, K. Bolotin, Y. Zhao, S. Adam, E.H. Hwang, S. Das Sarma,

H.L. Stormer, P. Kim, Measurement of scattering rate and minimum conductivity in graphene, Phys. Rev. Lett. 99 (2007) 246803. https://doi.org/10.1103/PhysRevLett.99.246803

[S15] C.T. White, J.W. Mintmire, Density of states reflects diameter in nanotubes, Nature 394 (1998) 29–30. https://doi.org/10.1038/27801

[S16] S.G. Lemay, J.W. Janssen, M. van den Hout, M. Mooij, M.J. Bronikowski, P.A. Willis, R.E. Smalley, L.P. Kouwenhoven, C. Dekker, Two-dimensional imaging of electronic wavefunctions in carbon nanotubes, Nature **412** (2001) 617–620. https://doi.org/10.1038/35088013

[S17] T. Ma, Z. Liu, J. Wen, Y. Gao, X. Ren, H. Chen, C. Jin, X.-L. Ma, N. Xu, H.-M. Cheng, W. Ren, Tailoring the thermal and electrical transport properties of graphene films by grain size engineering, Nat. Commun. **8** (2017) 14486. https://doi.org/10.1038/ncomms14486

[S18] M. Hayashi, Differential geometry and morphology of graphitic carbon materials, Phys. Lett. A **342** (2005) 237–246. https://doi.org/10.1016/j.physleta.2005.05.037